\documentclass[fp,twocolumn]{jpsj3}
\usepackage{txfonts}
\usepackage{graphicx,color}
\usepackage{braket}
\usepackage{bm}

\title{Temperature-driven and electrochemical-potential-driven adiabatic pumping \\
via a quantum dot}

\author{Masahiro Hasegawa and Takeo Kato}
\inst{Institute for Solid State Physics, The University of Tokyo, Kashiwa, Chiba 277-8581, Japan} 

\abst{
We investigate adiabatic pumping via a single level quantum dot induced by periodic modulation of thermodynamic variables of reservoirs, i.e., temperatures and electrochemical potentials.
We consider the impurity Anderson model and derive analytical formulas for coherent adiabatic charge pumping applicable to the strong dot-reservoir coupling within first-order perturbation with respect to Coulomb interaction.
We show that charge pumping is induced by rectification effect due to delayed response of the quantum dot to time-dependent reservoir parameters.
The presence of interaction is necessary because this delayed response rectifies charge current via Coulomb interaction.
For temperature-driven charge pumping, one-way pumping is realized regardless of reservoir temperatures when an energy level of the quantum dot locates near the Fermi level.
We clarify that this new feature of adiabatic pumping is caused by level broadening effect of the quantum dot due to strong dot-reservoir coupling.
}

\begin{document}
\maketitle

\section{Introduction}

Adiabatic pumping is a process by which a finite charge (and heat) current is induced under the periodic slow modulation of external control parameters.
Motivated by recent developments of experimental techniques, adiabatic pumping has been studied in a lot of theoretical and experimental papers.
For example, adiabatic charge pumping in nanoscale devices has been reported~\cite{Kouwenhoven91,Giazotto11,Connolly13,Roche13}, and has been applied to practical applications such as the current standard~\cite{Pekola13a}.
Adiabatic pumping has also been attracting attentions as a key element in study of nonequilibrium thermodynamics in small systems; for instance, heat and work exchange~\cite{Pekola13b,Gasparinetti14} and feedback control~\cite{Toyabe10,Sagawa12,Koski14} have been studied.

Investigation of a quantum nature in adiabatic pumping has been one of important issues for long time.
Theoretically, adiabatic pumping in mesoscopic systems under slow gate-voltage modulation was initially formulated for noninteracting systems by several researchers~\cite{Thouless83,Buttiker93,Buttiker94,Pretre98}, and then expressed in a simple formula by Brouwer~\cite{Brouwer98}.
Not only because effect of Coulomb interaction between electrons is important in a lot of mesoscopic systems, but also because nontrivial many-body effects such as the Kondo effect emerges, adiabatic pumping has also been theoretically studied for various interacting electron systems~\cite{Haupt13,Aleiner99,Citro03,Das05,Aono04,Splettstoesser05,Sela06,Fioretto08,Brouwer05,Splettstoesser06,Hernandez09,Reckermann10,Calvo12,Juergens13,Lim13,Nakajima15}.
All these theoretical works have investigated adiabatic pumping induced by gate voltage modulation because such pumping is realized rather easily in experiments.
However, other type of pumping, e.g. pumping induced by temperature modulation, is also important from a viewpoint of thermodynamics.
Adiabatic heat pumping induced by temperature modulation has first been discussed in phonon transport via molecule junctions~\cite{Ren10}, and subsequently, temperature-induced electronic transport has been discussed in thermoelectric transport~\cite{Juergens13,Simine12,Arrachea12} and non-equilibrium thermodynamics~\cite{Sagawa11,Yuge13}.
However, temperature-driven pumping has been considered only for an incoherent transport regime, where transport is described by Master equations on a basis of second-order perturbation with respect to a coupling strength between a system and reservoirs.
Although quantum effect, e.g. level broadening effect of the QD due to strong dot-reservoir coupling, becomes significant in adiabatic pumping at low temperatures, such coherent adiabatic pumping has not been formulated because of a lack of a systematic method to describe time-dependent reservoir temperatures in coherent transport.

\begin{figure}[tb]
\begin{center}
\includegraphics[clip,width=5.5cm]{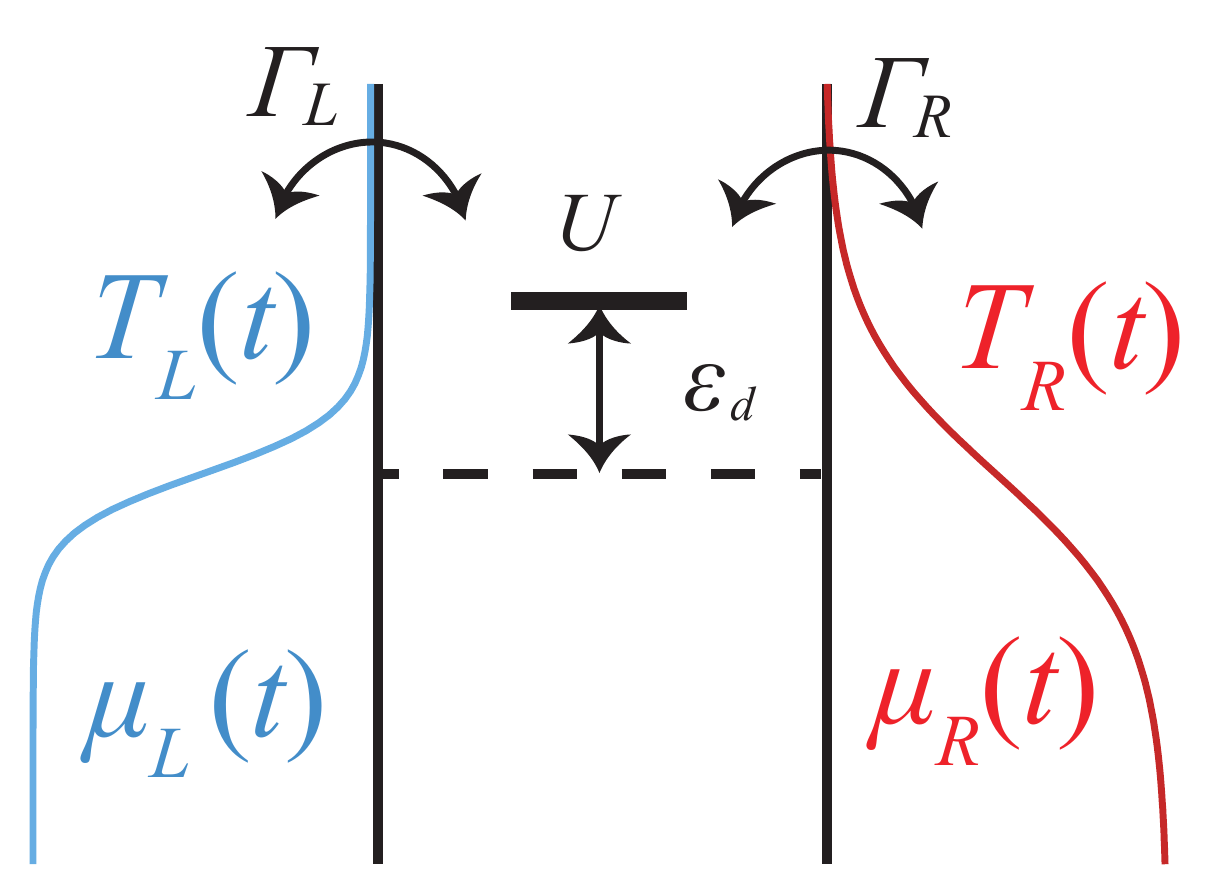}
\caption{\label{fig:dot_pic} 
A schematic figure of the system considered in this paper. 
A single-level QD is connected to two reservoirs, whose temperatures and electrochemical potentials are periodically modulated.}
\end{center}
\end{figure}

In this paper, we investigate adiabatic charge pumping via a single-level quantum dot (QD) coupled to two reservoirs with time-dependent temperatures and electrochemical potentials (as seen in Fig.~\ref{fig:dot_pic}) as the first step to construct non-equilibrium thermodynamics for coherent transport.
We consider the Anderson impurity model with Coulomb interaction as a simple setup for non-trivial coherent charge pumping driven by the thermodynamic variables of the reservoirs, and employ the Keldysh formalism~\cite{Jauho94}, which is applicable for arbitrary strength of dot-reservoir coupling.
To describe time-dependent temperature modulation in the Keldysh formalism, we employ the method of the thermomechanical field~\cite{Luttinger64,Eich14,Tatara15,footnote:tmfield}.

We first derive an analytical formula of adiabatic charge pumping within the first-order perturbation theory with respect to Coulomb interaction.
By this formula, we clarify that the Coulomb interaction in the QD is essential for adiabatic pumping induced by the reservoir parameters driving, and that no charge is pumped for non-interacting systems. 
We show that the present charge pumping is induced by rectification effect due to delayed response of an effective QD energy level with respect to time-dependent reservoir parameters.
The effective QD energy level is shifted via Coulomb interaction by the occupation of the QD which responses time-dependent reservoir parameters with delay time, and this delayed energy shift rectifies the charge current and produces net charge pumping.
The relation between charge pumping and effective energy shift of the QD is extension of that for incoherent transport discussed in some papers~\cite{Reckermann10,Calvo12}.
We also show that, for temperature-driven pumping, one-way pumping can be realized regardless of the reservoir temperatures when the QD energy level locates near the Fermi level.
This is a novel feature of adiabatic coherent pumping due to the level broadening effect of the QD.
Finally, we estimate pumping current in a realistic experimental setup.

This paper is organized as follows.
In Sect.~\ref{sec:formulation}, we define a model Hamiltonian and Green's functions (GFs); we also introduce a thermomechanical field, and give a brief explanation of how it modulates the reservoir temperatures.
The adiabatic approximation is introduced at the end of this section.
In Sect.~\ref{sec:result}, we derive a formula for a charge pumped in one cycle of reservoir parameter modulation, and discuss mechanism of the present charge pumping.
In Sect.~\ref{sec:Numerical_calculation}, we show numerical results of pumping strength for both the temperature-driven pumping and the electrochemical-potential-driven pumping, and clarify their properties in detail.
We discuss experimental relevance, and relates the delayed response of the QD with a linear AC response of the QD in Sect.~\ref{sec:discussion}, and summarize our results in Sect.~\ref{sec:summary}.
In the Appendices, we present some technical details for certain relevant equations.

\section{\label{sec:formulation}Formulation}

In this section, we explain our model and some calculation methods.
We study the impurity Anderson model as a simple model for adiabatic charge pumping, and analytically calculate a pumped charge up to the first order of a Coulomb interaction in the QD by perturbation theory in the Keldysh GF method. 
A time-dependent dot-reservoir coupling is introduced to describe time-dependent reservoir parameters (i.e. the temperatures and the electrochemical potentials of reservoirs).
We also apply the adiabatic approximation to the time-dependent parameters by assuming that their time variation is sufficiently slow.
Throughout this paper, we define a charge of electrons as $-e$ ($e>0$).

\subsection{\label{sec:model}Model}

We consider a single-level QD coupled to two reservoirs (Fig. \ref{fig:dot_pic}).
The corresponding Hamiltonian is given by
\begin{align}
	H(t) &= \sum_{r=L,R} \sum_{k,s} \epsilon_{r k} c^{\dagger}_{r ks} c_{r ks} + 
	\sum_{s} \epsilon_d d^{\dagger}_s d_s, \nonumber \\
	&+ \sum_{r=L,R}  \sum_{k,s} (v_{rk}(t) c^{\dagger}_{rks} d_s + \mathrm{h.c.}) 
	+ U d^{\dagger}_{\uparrow} d_{\uparrow} d^{\dagger}_{\downarrow} d_{\downarrow}.
\end{align}
The first and second terms describe the electron reservoirs and the QD, respectively, where $d^{\dagger}_s$ ($d_{s}$) is the creation (annihilation) operator of an electron in the QD with a spin $s =\{ \uparrow, \downarrow \}$, and $c^{\dagger}_{r ks}$ ($c_{r ks}$) is that of an electron in the reservoir $r$ ($=L, R$) with a wave number $k$ and a spin $s$.
The electron energies in the QD and the reservoirs are denoted by $\epsilon_d$ and $\epsilon_{rk}$, respectively.
The third term describes the dot-reservoir coupling, while the last term describes the Coulomb interaction between electrons in the QD, and $U$ denotes its strength.
We prepare the reservoirs to be initially in equilibrium and have the same temperature $T_{\rm ref}$ and the same electrochemical potential (the latter is set to zero in this paper).
We call $T_{\rm ref}$ the reference temperature hereafter.
In the present model, we have introduced a time-dependent dot-reservoir coupling $v_{rk}(t)$ to describe time-dependence of the reservoir parameters.
Further details are given in Sect.~\ref{sec:TemperatureModulation}.

\subsection{Keldysh Green's function}
\label{sec:gf}

A pumping current is formulated within the framework of the Keldysh GFs. 
The retarded, advanced, and lesser components of the Keldysh GFs for electrons in the QD for the noninteracting systems ($U=0$) are defined as
\begin{align}
	G_{0,s}^{R}(t,t^{\prime}) 
	&= -i \Theta(t - t^{\prime}) \Braket{[ d^{\dagger}_s(t^{\prime}) , d_s(t) ]_+ }_0, \\
	G_{0,s}^{A}(t,t^{\prime}) 
	&= i \Theta(t^{\prime} - t) \Braket{[ d^{\dagger}_s(t^{\prime}) , d_s(t) ]_+ }_0, \\
	G_{0,s}^{<}(t,t^{\prime}) &= i \Braket{ d^{\dagger}_s(t^{\prime}) d_s(t) }_0,
\end{align}
where $\Theta(t)$ is the Heaviside step function.
These components of the Keldysh GF are calculated using the Dyson equation as
\begin{align} \label{eqn:unperturbed_dyson_eqn}
	G_{0,s}^R(t,t^{\prime}) &= g_{d,s}^R(t,t^{\prime}) \nonumber \\
	&+ \int dt_1 dt_2 g_{d,s}^R(t,t_1) \Sigma^R_{s}(t_1,t_2) G_{0,s}^R(t_2,t^{\prime}), \\
	G_{0,s}^A(t,t^{\prime}) &= \bigl( G_{0,s}^R(t^{\prime},t) \bigr)^\ast, \\
	G_{0,s}^<(t,t^{\prime}) &= \int dt_1 dt_2 G_{0,s}^R (t,t_1) \Sigma_s^<(t_1,t_2) G_{0,s}^A(t_2,t^{\prime}),
\end{align}
where $g_d^{R}(t,t^{\prime})$ and $g_d^{A}(t,t^{\prime})$ are the retarded and advanced GFs for electrons in the isolated QD, respectively.
The retarded and lesser components of the self-energy, $\Sigma_s^R(t_1,t_2)$ and $\Sigma_s^<(t_1,t_2)$, are given by
\begin{align}
	\Sigma_s^R(t_1,t_2) &= \sum_{r=L,R} \sum_{k} v_{r k}^{\ast}(t_1) g_{rks}^R(t_1,t_2) v_{r k}(t_2), \label{eq:self1} \\
	\Sigma_s^<(t_1,t_2) &= \sum_{r=L,R}  \sum_{k} v_{r k}^{\ast}(t_1) g_{rks}^<(t_1,t_2) v_{r k}(t_2), \label{eq:self2}
\end{align}
where $g_{rks}^{R}(t,t^{\prime})$ and $g_{rks}^{<}(t,t^{\prime})$ are the retarded and lesser GFs for electrons in the isolated reservoir $r$, respectively.

The Keldysh GFs for the interacting electron system are defined in a similar manner:
\begin{align}
	G_{s}^{R}(t,t^{\prime}) &= -i \Theta(t - t^{\prime}) \Braket{[ d^{\dagger}_s(t^{\prime}) , d_s(t) ]_+ } , \\
	G_{s}^{A}(t,t^{\prime}) &= i \Theta(t^{\prime} - t) \Braket{[ d^{\dagger}_s(t^{\prime}) , d_s(t) ]_+ } , \\
	G_{s}^{<}(t,t^{\prime}) &= i \Braket{ d^{\dagger}_s(t^{\prime}) d_s(t) } .
\end{align}
In this paper, the charge current is evaluated up to the first order of $U$ by approximating the Keldysh GFs as follows:
\begin{align}
	G_s^R(t,t^{\prime}) &\simeq G_{0,s}^R(t,t^{\prime}) \nonumber \\
	&- iU \int dt_1 G_{0,s}^R(t,t_1) G_{0,\bar{s}}^<(t_1,t_1) G_{0,s}^R(t_1,t^{\prime}), \label{eqn:first_U_R} \\
	G_s^A(t,t^{\prime}) &\simeq G_{0,s}^A(t,t^{\prime}) \nonumber \\
	& - iU \int dt_1 G_{0,s}^A(t,t_1) G_{0,\bar{s}}^<(t_1,t_1) G_{0,s}^A(t_1,t^{\prime}). \label{eqn:first_U_A}
\end{align}
where $\bar{s}$ indicates the spin opposite to $s$.
The lesser GF is evaluated up to the first order of $U$ by substituting (\ref{eqn:first_U_R}) and (\ref{eqn:first_U_A}) into its definition given by
\begin{align}\label{eqn:first_U_<}
	G_{s}^{<}(t,t^{\prime}) = \int dt_1dt_2 G_s^R(t,t_1) \Sigma_s^<(t_1,t_2) G_s^A(t_2,t^{\prime}).
\end{align}

\subsection{\label{sec:c_c}Transferred charge}

The charge current flowing from the reservoir $r$ into the QD is given by
\begin{align} \label{eqn:c_cur_ori}
       J_{r}(t) = -2e \sum_{k,s} \operatorname{Re} \left[ v_{r ks}(t) G^<_{s;rks}(t,t) \right],
\end{align}
where $G^<_{s;rks}(t,t')$ is defined as
\begin{align}
       G^<_{s;rks}(t,t^{\prime}) = i \Braket{c_{rks}^{\dagger}(t^{\prime}) d_s(t)}.
\end{align}
The charge accumulated from the reservoir $r$ between the initial time $t_i$ and the finial time $t_f$ is
\begin{align}
       \delta Q_{r} = \int_{t_i}^{t_f} dt J_{r}(t).
       \label{eqn:time_int}
\end{align}
To consider the accumulated charge in a cycle, we set $t_f -t_i = \mathcal{T}$, where $\mathcal{T}$ is a period of of the time-dependent reservoir parameters.
Using the method of the equation of motion~\cite{Jauho94}, the pumped charge can be rewritten as
\begin{align}\label{eqn:acm_cur}
	\delta Q_{r} &= -2e \sum_{k,s} \int_{t_i}^{t_f} dt \int dt_1 \nonumber \\
	& \hspace{5mm} \times\mathrm{Re} \Bigl[ v_{rks}(t)G_{s}^R(t,t_1) v^{\ast}_{rks}(t_1)g_{rks}^<(t_1,t) \nonumber \\
	& \hspace{10mm} + v_{rks}(t)G_{s}^<(t,t_1) v^{\ast}_{rks}(t_1)g_{rks}^A(t_1,t) \Bigr].
\end{align}

\subsection{Time-dependent dot-reservoir coupling}
\label{sec:TemperatureModulation}

To treat time-dependent thermodynamic variables of the reservoirs, we introduce a method of the time-dependent dot-reservoir coupling, which is equivalent to the thermomechanical field method~\cite{Luttinger64,Eich14,Tatara15,footnote:tmfield}.
We consider the time-dependent dot-reservoir coupling given by
\begin{align}
 v_{rk}(t) = v_{rk} \sqrt{1+ \dot{B}_{r}(t)} e^{i (\epsilon_{rk} B_{r}(t) + \Lambda_{r}(t))},
\end{align}
where $B_r(t)$ and $\Lambda_r(t)$ are functions of time related with the temperature and the electrochemical potential of the reservoir $r$, respectively.
With this time-dependent dot-reservoir coupling, the self-energies (Eqs.~(\ref{eq:self1}) and (\ref{eq:self2})) become
\begin{align}\label{eqn:self_ene_R_gen}
	\Sigma_s^R(t_1,t_2) &= \sum_{r=L,R} \sum_{k} |v_{rk}|^2 g^{R}_{rks}(t_1,t_2)C_r(\epsilon_{rk},t_1,t_2), 
	 \\
	\Sigma_s^<(t_1,t_2) &= \sum_{r=L,R} \sum_{k} |v_{rk}|^2 g^{<}_{rks}(t_1,t_2)C_r(\epsilon_{rk},t_1,t_2), 
	\label{eqn:self_ene_R_gen2}\\
	C_{r}(\omega , t_1, t_2) &=  \sqrt{(1+\dot{B}_{r}(t_1))(1+\dot{B}_{r}(t_2)) } \nonumber \\
       &\hspace{5mm} \times e^{-i\omega (B_{r}(t_1)  - B_{r}(t_2) )
       - i(\Lambda_r(t_1) -\Lambda_r(t_2))}. \label{eq:Cdef}
\end{align}
Hereafter we assume the energy-independent dot-reservoir coupling constant ($v_{rk} = v_r$), and define the line width as $\Gamma_{r} = 2\pi \rho |v_{r}|^2$, where $\rho$ is a density of state at the Fermi level (the wide-band limit).

To highlight the effect of the time-dependent dot-reservoir coupling on the temperatures and the electrochemical potentials, let us first consider a simple example in which $B_r(t)$ and $\Lambda_r(t)$ have a linear time-dependence:
\begin{align}
	B_r(t) = \tilde{B}_r t, \quad \Lambda_r(t) = \tilde{\Lambda}_r t .
\end{align}
In this example, the retarded and lesser self-energies are given by
\begin{align}
	\Sigma_s^R(t_1,t_2) &= -i \sum_{r=L,R} \frac{\Gamma_{r}}{2} \delta(t_1 -t_2), \label{eq:TempModulatedSelfEnergy1} \\
	\Sigma_s^<(t_1,t_2) &= \int \frac{d\omega}{2\pi} \Sigma_s^<(\omega) e^{-i\omega ( t_1 - t_2) }, \\
	\Sigma_s^<(\omega) &= i \sum_{r=L,R} \Gamma_{r} f \left( \frac{\omega-\tilde{\Lambda}_r}{1+\tilde{B}_r} \right),
	\label{eq:TempModulatedSelfEnergy2}
\end{align}
where $f(\omega) =(e^{\omega/T_{\rm ref}} + 1)^{-1}$ is the Fermi distribution function at the reference temperature $T_{\rm ref}$ (refer to Appendix\ref{apx:selfenergy} for further details).
As seen in the expressions of the self-energies, the time-dependent dot-reservoir coupling solely affects the energy rescaling and the chemical-potential shift of the Fermi distribution function $f(\omega)$.
This modulation of the Fermi distribution function can be regarded as a change of the temperatures and the electrochemical potentials according to
\begin{align}
	\label{eq:relationT}
	T_r &= (1+\tilde{B}_r) T_{\rm ref}, \\
	\mu_r &= \tilde{\Lambda}_r.
	\label{eq:relationmu}
\end{align}
Thus, the present example describes the system in which the reservoir parameters are time-independent.

In general, $B_r(t)$ and $\Lambda_r(t)$ are not linear in time, and therefore, their effect to the reservoir parameters are non-trivial. 
However, when the modulation of $B_r(t)$ and $\Lambda_r(t)$ is sufficiently slow, the first time derivative of $B_r(t)$ and $\Lambda_r(t)$ can be related to the reservoir parameters in a similar way, as shown in the next subsection.

\subsection{\label{sec:ad_app}Adiabatic approximation}

Hereafter we consider the situation that the period $\mathcal{T}$ is much larger than a characteristic time scale of the system, which is $\Gamma^{-1} = (\Gamma_L + \Gamma_R)^{-1}$ in the present model.
In this situation, all the physical quantities are expected to be expanded with a small parameter $(\Gamma \mathcal{T})^{-1}$.
In order to expand the pumped charge given in Eq.~(\ref{eqn:acm_cur}) with respect to $(\Gamma \mathcal{T} )^{-1}$, let us consider a Tayler expansion of $B_r(t)$ and $\Lambda_r(t)$,
\begin{align}
	B_r(t_1) &= B_r(t) + \dot{B}_r(t)(t_1-t) + \ddot{B}_r(t_1) \frac{(t_1-t)^2}{2} + \cdots, \label{eq:BrExpansion}\\
	\Lambda_r(t_1) &= \Lambda_r(t) + \dot{\Lambda}_r(t)(t_1-t) + \ddot{\Lambda}_r(t) \frac{(t_1-t)^2}{2}+ \cdots,
	\label{eq:LambdaExpansion}
\end{align}
where $t_1$ is an integral variable, and $t \in [t_i,t_f]$ is a time variable of the current.
This Tayler expansion is a power series of $\mathcal{T}^{-1}$, since $\dot{B}_r(t) \sim \mathcal{T}^{-1} B_r(t)$, $\ddot{B}_r(t) \sim \mathcal{T}^{-2} B_r(t)$, $\cdots$ hold.
This expansion may cause confusion to the reader because $B_r(t)$ and $\Lambda_r(t)$ are expanded not with $(\Gamma \mathcal{T} )^{-1}$ but with $\mathcal{T}^{-1}$.
However, this expansion actually leads to a power series of observables with respect to $(\Gamma \mathcal{T})^{-1}$ after integration with respect to internal time variables as shown at the end of this subsection.

To evaluate the contribution of the linear term, i.e. the first time derivatives of $B_r(t)$ and $\Lambda_r(t)$, let us first consider the first-order approximation in Eqs.~(\ref{eq:BrExpansion}) and (\ref{eq:LambdaExpansion}).
By calculating the self-energies given in Eqs.~(\ref{eqn:self_ene_R_gen})-(\ref{eq:Cdef}), the time derivatives of $B_r(t)$ and $\Lambda_r(t)$ are related to the time-dependent reservoir parameters as
\begin{align} \label{eqn:externalfeld}
	&T_r(t) = (1+\dot{B}_r(t)) T_{\rm ref}, \\
	& \mu_r(t) = \dot{\Lambda}_r(t), \label{eqn:externalfeld2}
\end{align}
where $T_r(t)$ and $\mu_r(t)$ is the temperatures and the electrochemical potentials of the reservoir $r$ at time $t$. 
In this derivation, we have assumed the condition 
\begin{align}
1 + \dot{B}_{r}(t) > 0, \label{eq:positiveness}
\end{align}
to guarantee the positiveness of the temperature.
Note that these relations can be regarded as an extension of Eqs.~(\ref{eq:relationT}) and (\ref{eq:relationmu}) for the steady state under time-independent reservoir parameters.

Let us now introduce the adiabatic approximation.
The adiabatic approximation considers expansion up to the first order of $\dot{T}_r(t)$ and $\dot{\mu}_r(t)$ and ignores other non-adiabatic terms, e.g. the terms proportional to $\ddot{T}_r(t)$ and $(\dot{T}_r(t))^2$.
In the present formalism, this approximation is equivalent to the second-order approximation, i.e., the approximation up to the second-order time derivatives in the Tayler expansions, Eqs.~(\ref{eq:BrExpansion}) and (\ref{eq:LambdaExpansion}).
Since the function $C_r$ (see Eqs.~(\ref{eqn:self_ene_R_gen})-(\ref{eq:Cdef})) contains all the time-dependences of $B_r(t)$ and $\Lambda_r(t)$, it is sufficient to expand it for the adiabatic approximation.
By expanding $C_{r}(\omega , t_1+t, t_2+t)$ with respect to $t_1$ and $t_2$, one can obtain
\begin{align}
      & C_{r}(\omega , t_1+t, t_2+t) \simeq C_{r}^{(0)}(\omega , t_1, t_2; t) +  C_{r}^{(1)}(\omega , t_1, t_2;t), \\
      & C_{r}^{(0)}(\omega ,t_1, t_2;t) \nonumber \\
      & \quad = (1+\dot{B}_{r}(t)) e^{-i [\omega \dot{B}_{r}(t) +\dot{\Lambda}_{r}(t)] (t_1 -t_2)} ,\label{eqn:ad_c_0} \\
      & C_{r}^{(1)}(\omega , t_1, t_2;t)  \nonumber \\
      & \quad = \Biggl\{ \frac{\ddot{B}_{r}(t)}{2} \left[ (t_1 +t_2 ) + i\omega (1+\dot{B_{r}}(t)) (t_2^2 - t_1^2) \right] \nonumber \\
      & \qquad + \frac{ i \ddot{\Lambda}_{r}(t)}{2} (t_2^2 - t_1^2)(1+\dot{B_{r}}(t)) \Biggr\} 
      e^{-i [\omega \dot{B}_{r}(t) +\dot{\Lambda}_{r}(t)] (t_1 -t_2)}. \label{eqn:ad_c_1}
\end{align}
where $C^{(0)}_{r}(\omega , t_1, t_2;t)$ is a steady-state contribution and $C^{(1)}_{r}(\omega , t_1, t_2;t)$ is a correction which contains the second-order time derivatives of $B_r(t)$ and $\Lambda_r(t)$.
Here $t \in [t_i , t_f]$ is a time variable of the current, and $t_1,t_2 \in [-\infty, \infty]$ are integral variables.

In this approximation, the lesser component of the self-energy becomes
\begin{align}\label{eqn:sigma_app}
	\Sigma^<(t_1,t_2) \simeq \Sigma^{<,(0)}(t_1,t_2;t) + \Sigma^{<,(1)}(t_1,t_2;t),
\end{align}
where $\Sigma^{<,(0)}(t_1,t_2;t)$ and $\Sigma^{<,(1)}(t_1,t_2;t)$ are proportional to $C^{(0)}_{r}(\omega , t, t_1, t_2)$ and $C^{(1)}_{r}(\omega , t, t_1, t_2)$, respectively.
The lowest order term, $\Sigma^{<,(0)}(t_1,t_2;t)$, describes the steady-state contribution for fixed values of the temperatures $T_r(t) = (1+\dot{B}_r(t))T_{\rm{ref}}$ and the electrochemical potentials $\mu_r(t) = \dot{\Lambda}_r(t)$.
The next term $\Sigma^{<,(1)}(t_1,t_2;t)$, which is proportional to time derivatives of reservoir parameters, i.e., $\dot{T}_r(t) = \ddot{B}_r(t) T_{\rm{ref}}$ and $\dot{\mu}_r(t) = \ddot{\Lambda}_r(t)$, describes the correction to the steady-state contribution.
Thus, by the adiabatic approximation, the pumping current can be evaluated as a sum of the steady-state contribution and the correction to it up to the first order of $\dot{T}_r(t)$ and $\dot{\mu}_r(t)$.

As shown in Eq.~(\ref{eqn:sigma_app}), the Tayler expansion of $B_r(t)$ and $\Lambda_r(t)$ is equivalent with the expansion of the self-energies.
In the quantum field theory, any observables are expressed by convolution time integral of the GFs and the self-energies.
In the present model, the noninteracting GFs decays with a decay rate $\Gamma$, i.e. $G_{0,s}^{R}(t,t^{\prime}) \sim e^{-\Gamma |t-t^{\prime}| }$, and time integral of such GFs produces the factor $\Gamma^{-1}$.
Thus, once the self-energies are expanded with respect to $\mathcal{T}^{-1}$, then time integration of the GFs automatically gives the expansion of the observables with respect to $(\Gamma \mathcal{T})^{-1}$.

\section{\label{sec:result}Analytical Result}

\subsection{Formula for the pumped charge}

The transferred charge given by Eq.~(\ref{eqn:acm_cur}) is calculated up to the first order of $U$ by substituting Eqs.~(\ref{eqn:first_U_R}), (\ref{eqn:first_U_A}) and (\ref{eqn:first_U_<}) into (\ref{eqn:acm_cur}).
The amount of the transferred charge from the reservoir $r$ in one period can be decomposed into a noninteracting term $\delta Q_{r,0}$ and the first-order term $\delta Q_{r,U}$ proportional to $U$ as follows:
\begin{align}
	\delta Q_{r} = \delta Q_{r,0} + \delta Q_{r,U} + O(U^2).
\end{align}
The explicit forms of $\delta Q_{r,0}$ and $\delta Q_{r,U}$ are presented in Appendix\ref{apx:rep_of_Q0Q1}.
The adiabatic transferred charge is obtained by applying the adiabatic approximation to $\delta Q_{r,0}$ and 
$\delta Q_{r,U}$ as 
\begin{align}
	\delta Q_{r,0} &= \delta Q_{r,0}^{(0)} + \delta Q_{r,0}^{\mathrm{pump}} + O((\Gamma \mathcal{T})^{-1}), \\
	\delta Q_{r,U} &= \delta Q_{r,U}^{(0)} + \delta Q_{r,U}^{\mathrm{pump}} + O((\Gamma \mathcal{T})^{-1}).
\end{align}
Here, $\delta Q_{r,0}^{(0)}$ and $\delta Q_{r,U}^{(0)}$ are steady-state contributions, which are calculated by approximating the self-energy up to the lowest order as $\Sigma^{<} \simeq \Sigma^{<,(0)}$, while $\delta Q_{r,0}^{\mathrm{pump}}$ and $\delta Q_{r,U}^{\mathrm{pump}}$ are the pumping contributions, which are proportional to the contribution of $\Sigma^{<,(1)} \sim O(\dot{T}_r,\dot{\mu}_r)$.
We note that the steady-state contributions are finite in general.
For example, when the electrochemical potential of one reservoir is always larger than that of the other reservoir throughout a cycle while keeping the temperatures constant, the steady-current contribution remains finite.
However, the pumping contributions of the transferred charge is distinguished from the steady-state one by the dependence on the pumping frequency $\mathcal{T}^{-1}$; the pumping contributions are independent of $\mathcal{T}$, whereas the steady-state contributions are proportional to $\mathcal{T}$.
Hereafter, we focus only on the pumping contributions.

In Sect.~\ref{sec:NoninteractingPump} and Sect.~\ref{sec:InteractingPump}, we derive formulas for $\delta Q_{r,0}^{\mathrm{pump}}$ and $\delta Q_{r,U}^{\mathrm{pump}}$ in a line-integral form.
Here we define some functions used in those formulas in advance.
We define a parameter vector by
\begin{align}\label{eqn:param_vec}
	\bm{X}(t) &= (T_L(t),T_R(t),\mu_L(t),\mu_R(t)) \nonumber \\
	&=(X^{\mu}(t) ), \quad (\mu=1,2,3,4),
\end{align}
where the vector components are defined by Eqs.~(\ref{eqn:externalfeld}) and (\ref{eqn:externalfeld2}).
We define a spectral function of the QD and a Fermi distribution function of the reservoir $r$ as
\begin{align}\label{eqn:spectralfunction}
	\mathcal{A}(\omega) &= \frac{1}{2\pi}\frac{\Gamma}{(\omega - \epsilon_d)^2 + \Gamma^2/4}, \\
	f_{r}(\omega; \bm{X}) &= f(\omega;T_r,\mu_r) = {[e^{(\omega -\mu_r)/ T_{r}} +1]}^{-1}, \label{eq:distributionfunc}
\end{align}
respectively. 
Using the Fermi distribution functions, we also define an effective distribution function of the QD as
\begin{align}
	f_{\mathrm{eff}}(\omega; \bm{X}) &= \frac{\Gamma_L f_L(\omega; \bm{X}) + \Gamma_R f_R(\omega; \bm{X})}{\Gamma}. \label{eqn:EffectiveFermi}
\end{align}

\subsubsection{A noninteracting part of the pumped charge}
\label{sec:NoninteractingPump}

First, we derive a formula for the noninteracting term $\delta Q_{r,0}^{\mathrm{pump}}$, and show that this term is canceled out in a period.
By substituting Eqs.~(\ref{eqn:ad_c_0}) and (\ref{eqn:ad_c_1}) into the expression of $\delta Q_{r,0}^{\rm pump}$ given in Eqs.~(\ref{eqn:noninteracting_tot})-(\ref{eqn:noninteracting_B}), $\delta Q_{r,0}^{\mathrm{pump}}$ becomes, 
\begin{align}
	\delta Q_{r,0}^{\mathrm{pump}} &= \sum_{\mu = 1}^{4}
	\int_{X_{\mu}(t_i)}^{X_{\mu}(t_f)} \! \! \! dX_{\mu} \pi_{r,0}^{\mu}(\bm{X}), \label{eqn:Q0pumpdef} \\
	\pi_{r,0}^{\mu}(\bm{X}) &= \frac{\partial}{\partial X_\mu} \biggl( 
	 e  \int d\omega \Gamma_r 
	 \bigl[ (\partial_{\omega} \mathcal{A}(\omega)) f_r(\omega; \bm{X}) \Biggr. \nonumber \\
	& \hspace{10mm} \biggl.
	- \pi \mathcal{A}^2(\omega) f_{\mathrm{eff}}(\omega; \bm{X}) \bigr]\biggr) , \label{eqn:Q0pump}
\end{align}
where $\sum_\mu \int dX_{\mu}$ is a line integral along the trajectory of $\bm{X}(t)$.
A detailed derivation is given in Appendix\ref{noninteracting_part}.
Since $\pi_{r,0}^{\mu}(\bm{X})$ is a gradient of a scalar function of $\bm{X}$, the noninteracting part always becomes zero ($\delta Q_{r,0}^{\mathrm{pump}} =0$) under the periodic boundary condition $\bm{X}(t_i) = \bm{X}(t_f)$.
Accordingly, no charge pumping is induced for a noninteracting electron model ($U=0$) by the time-dependent modulation of the reservoir parameters.
Hereafter, we drop $\pi_{r,0}^{\mu}(\bm{X})$ for the discussion of the charge pumping.

\subsubsection{An interacting part of the pumped charge}
\label{sec:InteractingPump}

Next, we derive a formula for the interacting part $\delta Q_{r,U}^{\mathrm{pump}}$.
By substituting Eqs.~(\ref{eqn:ad_c_0}) and (\ref{eqn:ad_c_1}) into $\delta Q_{r,U}^{\mathrm{pump}}$, 
it is obtained as
\begin{align}
	& \delta Q_{r,U}^{\mathrm{pump}} = 
	\sum_{\mu=1}^{4} \int_{\bm{X}(t_i)}^{\bm{X}(t_f)} dX_{\mu} \pi_{r,U}^{\mu}(\bm{X}), 
	\label{eqn:gene_pump_c} \\
	&\pi_{r,U}^{\mu}(\bm{X}) = \pi_{r,U,\mathrm{non-pump}}^{\mu}(\bm{X}) + \pi_{r,U,\mathrm{pump}}^{\mu}(\bm{X})  \label{eqn:addpump} \\
	&\pi_{r,U,\mathrm{non-pump}}^{\mu}(\bm{X}) \nonumber \\
	&= -\frac{\partial}{\partial X_{\mu}} \biggl\{ \frac{2\pi e U}{\Gamma^3} \int d\omega_1 \mathcal{A}(\omega_1) f_{\mathrm{eff}}(\omega_1;\bm{X}) \nonumber \\
	& \qquad \times \int d\omega_2 \sum_{r_1}  \Bigl[ \Gamma \delta_{r,r_1} \frac{\partial }{\partial \omega_2} \left( \frac{1}{\pi \Gamma}\mathcal{A}(\omega_2) - \mathcal{A}^2(\omega_2) \right) \nonumber \\
	& \qquad \qquad \qquad \qquad + \Gamma_r  \frac{\partial \mathcal{A}^2(\omega_2) }{\partial \omega_2} \Bigr] \Gamma_{r_1} f_{r_1}(\omega_2;\bm{X}) \biggr\}  \\
	&\pi_{r,U,\mathrm{pump}}^{\mu}(\bm{X}) \nonumber \\
	&=-\frac{2\pi e U \Gamma_{r} \Gamma_{\bar{r}}}{\Gamma} \int d\omega_1 \frac{\partial \mathcal{A}(\omega_1)}{\partial \omega_1}  (f_{r}(\omega_1;\bm{X}) - f_{\bar{r}}(\omega_1;\bm{X}) ) \nonumber \\
	&\qquad \times \int d\omega_2 \mathcal{A}^2(\omega_2) \frac{\partial f_{\mathrm{eff}}(\omega_2;\bm{X})}{\partial X^{\mu}} . \label{eqn:pi_only_pump}
\end{align}
A detailed derivation is given in Appendix\ref{rep_of_pi}.
Contrary to the noninteracting term, $\pi_{r,U,\mathrm{pump}}^{\mu}(\bm{X})$ is not a gradient of a scalar function, and therefore, becomes finite under the time-dependent driving of the reservoir parameters.
We note that the non-pumping term, $\pi_{r,U,\mathrm{non-pump}}^{\mu}(\bm{X})$, does not contribute to the pumped charge under the periodic boundary condition $\bm{X}(t_i) = \bm{X}(t_f)$ since this term is given by a gradient of a scalar function of $\bm{X}$.
Hereafter, we drop $\pi_{r,U,\mathrm{non-pump}}^{\mu}(\bm{X})$ for the discussion of the charge pumping.

\subsection{Mechanism of charge pumping}
\label{sec:mechanism}

In this section, we explain physical mechanism of the present pumping by considering rectification effect induced by a delayed response of an effective QD energy level.


First, we discuss the delayed response of the effective QD energy level.
Considering the Hartree approximation, the effective QD energy level can be written up to the first order of $U$ as
\begin{align}\label{eqn:energyshift}
	\tilde{\epsilon}_{d,s}(t) = \epsilon_d + U \langle n_{\bar{s}}(t) \rangle_0 + O(U^2), 
\end{align}
where $\langle n_{s}(t) \rangle_0 = -iG_{0,s}^{<}(t,t)$ is the average occupation number of the noninteracting QD ($U=0$).
Using the adiabatic approximation, $\langle n_{s}(t) \rangle_0$ becomes
\begin{align}
	\langle n_{s}(t) \rangle_0 &= \langle n_{s}(t) \rangle_0^{(0)} + \langle n_{s}(t) \rangle_0^{(1)} + O((\Gamma \mathcal{T})^{-2}), 
	\label{eq:particlenumer}\\
	 \langle n_{s}(t) \rangle^{(0)}_0 &= \int d\omega \mathcal{A}(\omega) f_{\rm eff}(\omega;\bm{X}(t)), \\
	 \langle n_{s}(t) \rangle^{(1)}_0 &= - \sum_{\mu=1}^{4}
	 \int d\omega \pi \mathcal{A}(\omega)^2 \frac{\partial f_{\rm eff}(\omega;\bm{X}(t))}{\partial X^\mu} 
	 \dot{X}^\mu(t).
	 \label{eq:particlenumer2}
\end{align}
The leading term $\langle n_{s}(t) \rangle^{(0)}_0$ is the average occupation number with fixed reservoir parameters, $X^{\mu}(t)$, whereas the first-order term $\langle n_{s}(t) \rangle^{(1)}_0$ is a correction proportional to $\dot{X}^\mu(t)$.
To understand the effect of the correction term $\langle n_{s}(t) \rangle^{(1)}_0$ in terms of time delay, we rewrite it as
\begin{align}
\langle n_{s}(t) \rangle_0^{(1)} &= \langle n_{s}(t-\delta t) \rangle_0^{(0)} - \langle n_{s}(t) \rangle_0^{(0)} + O((\Gamma \mathcal{T})^{-2}), \\
\delta t &= \frac{\displaystyle{\sum_{\mu = 1}^4 \int d\omega \pi \mathcal{A}^2(\omega) \frac{\partial f_{\mathrm{eff}}(\omega)}{\partial X^{\mu}} \dot{X}^{\mu} }}{ \displaystyle{\sum_{\mu = 1}^4\int d\omega \mathcal{A}(\omega) \frac{\partial f_{\mathrm{eff}}(\omega)}{\partial X^{\mu}} \dot{X}^{\mu}}}. \label{eq:deltat}
\end{align}
Here, we have introduced a delay time $\delta t$ by assuming that there is a delay when the occupation number $\langle n_{s}(t) \rangle^{(0)}_0$ changes in response to the time-dependent reservoir parameters. 
Using this time delay, we define the steady-state contribution of the effective QD energy level and its change during the time delay as
\begin{align}
	\tilde{\epsilon}_d^{(0)}(t) &= \epsilon_d + U \langle n_{\bar{s}}(t) \rangle_0^{(0)}, \\
	\delta \tilde{\epsilon}_d^{(0)}(t) &= \tilde{\epsilon}_d^{(0)}(t-\delta t) -\tilde{\epsilon}_d^{(0)}(t) \nonumber \\
	&= U \left( \langle n_{\bar{s}}(t-\delta t) \rangle_0^{(0)} - \langle n_{\bar{s}}(t)\rangle_0^{(0)} \right) \nonumber \\
	&= U \langle n_{\bar{s}}(t) \rangle_0^{(1)} + O((\Gamma \mathcal{T})^{-2}), \label{eqn:delayQDenergy}
\end{align}
respectively. 
Then, the pumped charge given in Eqs.~(\ref{eqn:gene_pump_c})-(\ref{eqn:pi_only_pump}) is rewritten as
\begin{align}
	\delta Q_{r,U}^{\mathrm{pump}} &= \int_{t_i}^{t_f} dt \frac{\partial J_r^{\mathrm{steady}}(\epsilon_d,\bm{X}(t))}
	{\partial \epsilon_d} \delta \tilde{\epsilon}_d^{(0)}(t)  \nonumber \\
	& \hspace{5mm} +O(U^2) + O((\Gamma \mathcal{T})^{-1}), \label{eq:pumpcharge2} 
\end{align}
where $J^{\mathrm{steady}}_{r}(\epsilon_d,\bm{X})$ is a steady-state charge current with a fixed parameter $\bm{X}(t)$ given by
\begin{align}
	J^{\mathrm{steady}}_{r}(\epsilon_d, \bm{X}) &=- \frac{ 2 e\Gamma_{r} \Gamma_{\bar{r}}}{\Gamma} \int d\omega \mathcal{A}(\omega; \epsilon_d)
	 \nonumber \\& \hspace{8mm} \times 
	 \left[ f_r(\omega; \bm{X}) - f_{\bar{r}}(\omega; \bm{X}) \right]. \label{eqn:steady_current} 
\end{align}
Here, we have denoted the spectral function with $\mathcal{A}(\omega; \epsilon_d)$ to show its $\epsilon_d$-dependence explicitly (see Eq.~(\ref{eqn:spectralfunction})).
As seen in Eq.~(\ref{eq:pumpcharge2}), the pumped charge is directly induced by the energy shift of the effective QD energy level due to the delayed response of the occupation number in the QD.
We note that no charge pumping is induced in a noninteracting model because the effective QD energy level is constant ($\tilde{\epsilon}_d^{(0)}(t) = \epsilon_d$), and does not reflect the delayed response of the occupation number in the QD.

\begin{figure}[tb]
\begin{center}
\includegraphics[clip,width=7cm]{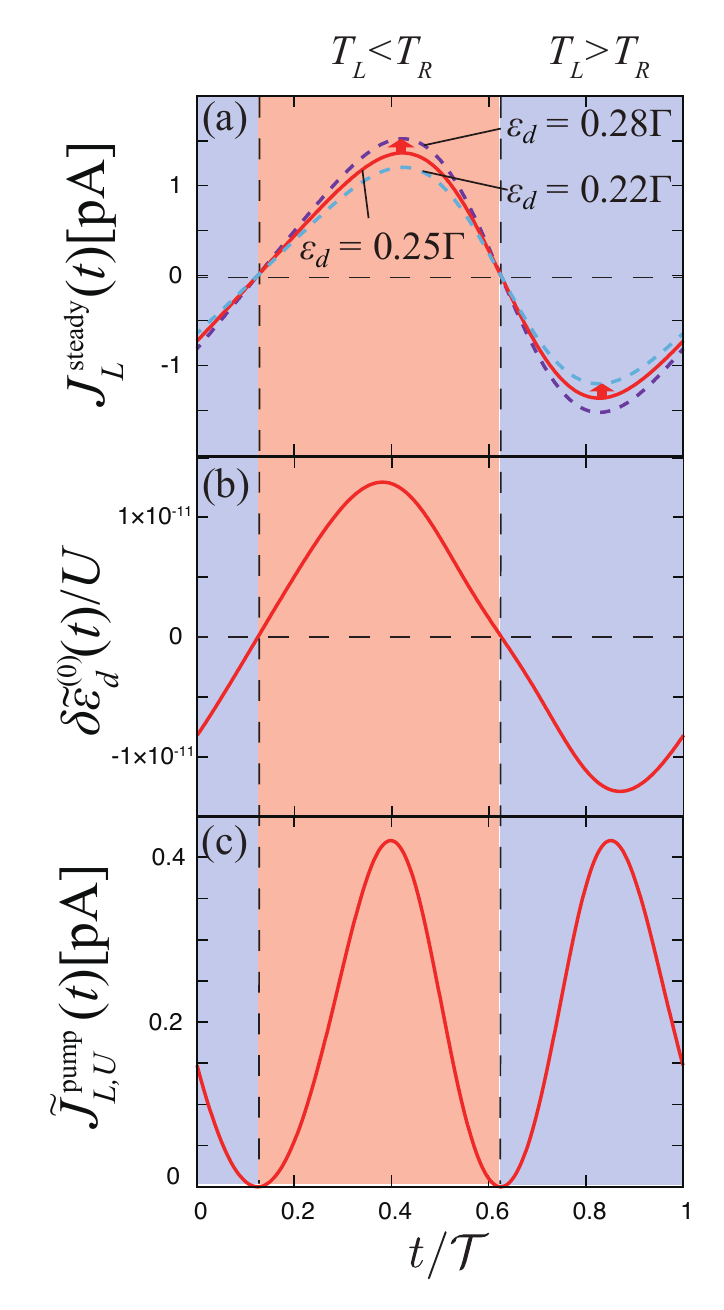}
\end{center}
\caption{\label{fig:timeplot_temp} Time-dependence of (a) the steady charge current, (b) the energy shift of the effective QD energy level due to time delay, and (c) the interacting part of the normalized pumping current. The temperatures are taken as $(T_L , T_R) = (\Gamma +  0.3\Gamma \cos (2\pi t/\mathcal{T}), \Gamma + 0.3\Gamma \sin (2\pi t/\mathcal{T}))$, and a symmetric coupling ($\Gamma_L =\Gamma_R=\Gamma/2$) is assumed.
The other parameters are set as $\epsilon_d = 0.25 \Gamma$ and $\mathcal{T}^{-1} =1 \mathrm{GHz}$.
In order to indicate the sign of $\partial_{\epsilon_d} J^{\mathrm{steady}}_{L}$, the steady charge currents for $\epsilon_d = 0.22 \Gamma$ and $0.28 \Gamma$ are also shown by the dashed lines.
In the region of $T_L < T_R$, the steady current is positive, and its magnitude is enhanced by the time delay of the effective QD energy level, because $\partial_{\epsilon_d} J^{\mathrm{steady}}_{L}$ is positive and the effective QD energy level become larger due to the time delay.
Conversely, in the region of $T_L > T_R$, the steady current is negative, and its magnitude is suppressed by the time delay, because $\partial_{\epsilon_d} J^{\mathrm{steady}}_{L}$ is negative and  the effective QD energy level become smaller due to the time delay.
As a result of this rectification effect, the positive pumping current is induced in both of the regions.}
\end{figure}

Next, we explain how the pumping current is induced by the energy shift of the effective QD energy level due to time delay.
For demonstration, let us consider the temperature-driven pumping by setting the reservoir parameters as $(T_L(t),T_R(t)) = (\Gamma + 0.3\Gamma \cos (2\pi \mathcal{T}^{-1} t),\Gamma + 0.3\Gamma \sin (2\pi \mathcal{T}^{-1} t))$ and $\mu_L = \mu_R = 0$.
Furthermore, we set the model parameters as $\Gamma_L = \Gamma_R = \Gamma/2$ and $\epsilon_d = 0.25 \Gamma$.
In Fig.~\ref{fig:timeplot_temp}~(a) and (b), we show time dependence of the steady charge current $J^{\mathrm{steady}}_{L}(\epsilon_d,\bm{X}(t))$ and the energy shift of the effective QD energy level due to time delay, $\delta \tilde{\epsilon}_d^{(0)}(t)$, respectively. 
In Fig.~\ref{fig:timeplot_temp}~(c), we also show the normalized pumping current $\tilde{J}_{L,U}^{\mathrm{pump}}(t)$ defined as
\begin{align}
	\tilde{J}_{L,U}^{\mathrm{pump}}(t) &= \frac{\Gamma}{U} J_{L,U}^{\mathrm{pump}}(t) \\
	J_{L,U}^{\mathrm{pump}}(t) &\equiv \sum_{\mu=1}^{4} \pi_{L,U,\mathrm{pump}}^{\mu}(\bm{X}(t)) \dot{X}^{\mu}(t) \nonumber \\
	&\hspace{-5mm} =\frac{\partial J_r^{\mathrm{steady}}(\epsilon_d,\bm{X}(t))}
	{\partial \epsilon_d} \delta \tilde{\epsilon}_d^{(0)}(t)+O(U^2) + O((\Gamma \mathcal{T})^{-1}). \label{eq:defpumpcurrent}
\end{align}
As indicated in Fig.~\ref{fig:timeplot_temp}, there are two time regions, i.e., the region of $T_L(t) < T_R(t)$ ($\mathcal{T}/8 < t < 5\mathcal{T}/8$) and that of $T_L(t) > T_R(t)$ ($0< t < \mathcal{T}/8, 5\mathcal{T}/8 < t < \mathcal{T} $).
Let us first consider the time region of $T_L(t) < T_R(t)$.
In this region, the steady charge current $J^{\mathrm{steady}}_{L}(\epsilon_d,\bm{X}(t))$ is positive, since its sign depends on the direction of the temperature bias.
At the same time, $\partial_{\epsilon_d} J^{\mathrm{steady}}_{L}(\epsilon_d,{\bm X}(t))$ is positive in this region as indicated by the two dashed lines drawn for $\epsilon_d = 0.22\Gamma$ and $0.28\Gamma$ in Fig.~\ref{fig:timeplot_temp}~(a).
On the other hand, the occupation number of the QD, $\langle n_{s}(t) \rangle^{(0)}_0$, takes a maximum value at $t = \mathcal{T}/8$ and a minimum value at $t = 5\mathcal{T}/8$ within the cycle for the symmetric coupling ($\Gamma_L = \Gamma_R$), and therefore, the occupation number of the QD continues to decrease in the region of $T_L(t) < T_R(t)$.
Then, the time delay induces the positive energy shift (i.e., $\delta \tilde{\epsilon}_d^{(0)}(t) > 0$) via the Hartree correction since the occupation number of the QD at time $t-\delta t$ is larger than at time $t$.
By combining this result with the sign of $\partial_{\epsilon_d} J^{\mathrm{steady}}_{L}(\epsilon_d,{\bm X}(t))$, we can conclude that the delayed response of the effective QD energy level enhances magnitude of the charge current when the steady current is positive.
By a similar discussion for the time region of $T_L(t) > T_R(t)$, we can also conclude that effect of the time delay suppresses magnitude of the charge current when the steady current is negative.
As a result, in both of the regions, the pumping current becomes positive and the net charge transferred in one cycle becomes finite.
Thus, the present mechanism of the charge pumping is regarded as `rectification' effect induced by the time delay.

In general, the direction of the rectification, i.e., the sign of the product of $\partial_{{\epsilon}_d} J^{\mathrm{steady}}_{L}({\epsilon}_d,{\bm X}(t))$ and $\delta \tilde{\epsilon}_{d}^{0)}(t)$, may change depending on the model parameters as discussed in Sect.~\ref{sec:numerical}.
Furthermore, the time dependence of the occupation number is nontrivial for the asymmetric coupling  ($\Gamma_L \ne \Gamma_R$).
However, the idea that the energy delay of effective QD energy level rectifies the charge current is still useful to understand the charge pumping for such a general case.

\subsection{Kernel representation}
\label{sec:kernelrep}

In this section, we introduce integral kernel representation for the pumped charge.
Using the Stokes' theorem, the line integral (Eq.~(\ref{eqn:gene_pump_c})) is rewritten into a surface integral.
This representation is convenient for visualization of the charge pumping strength discussed in Sect.~\ref{sec:Numerical_calculation}.

First, we show a formula for the pumped charge induced by the temperature driving assuming $\mu_L(t) = \mu_R(t) = 0$
The pumped charge is rewritten as
\begin{align}\label{eqn:pump_tempe}
	\delta Q_{L,U}^{\mathrm{pump}} &= \int_A dT_L dT_R 
	\left[ \frac{\partial \pi_L^2(\bm{X})}{\partial T_L} - \frac{\partial \pi_L^1(\bm{X})}{\partial T_R} \right] \nonumber \\
	&= \int_A dT_L dT_R \Pi_{T}(T_L,T_R),
\end{align}
where $A$ is an area whose boundary is the parameter trajectory $(T_L(t) , T_R(t))$.
The kernel $\Pi_T(T_L,T_R)$ is calculated by
\begin{align}
	\Pi_T(T_L,T_R) &= -\frac{2\pi eU \Gamma_{L} \Gamma_{R}}{\Gamma^2} \int d\omega d\omega^{\prime}
		 \left[  \partial_{\omega} \mathcal{A}(\omega) \mathcal{A}^2(\omega^{\prime}) \right. \nonumber \\
	&\hspace{-10mm} \left. \times \left( \Gamma_L \frac{\partial f_{L}(\omega^{\prime})}{\partial T_L} 
	\frac{\partial f_{R}(\omega)}{\partial T_R} + \Gamma_R \frac{\partial f_{L}(\omega)}{\partial T_L} 
	\frac{\partial f_{R}(\omega^{\prime})}{\partial T_R} \right) \right]. \label{eqn:integral_kernel_T}
\end{align}

Next, we show a formula for the pumped charge induced by the electrochemical-potential driving assuming $T_L(t) =T_R(t) = \mathrm{const.}$
The pumped charge is rewritten as
\begin{align}\label{eqn:pump_volt}
	\delta Q_{L,U}^{\mathrm{pump}} &= \int_{A^{\prime}} d\mu_L d\mu_R 
	\left[  \frac{\partial \pi_L^4(\bm{X})}{\partial \mu_L} - \frac{\partial \pi_L^3(\bm{X})}{\partial \mu_R}\right] \nonumber \\
	&= \int_{A^{\prime}} d\mu_L d\mu_R \Pi_{V}(\mu_L,\mu_R), 
\end{align}
where $A^{\prime}$ is an area whose boundary is the parameter trajectory $(\mu_L(t),\mu_R(t))$.
The kernel $\Pi_V(\mu_L,\mu_R)$ is calculated by
\begin{align}
	\Pi_V(\mu_L,\mu_R) &= -\frac{2\pi eU \Gamma_L \Gamma_R}{\Gamma^2} \int d\omega d\omega^{\prime} 
	\partial_{\omega} \mathcal{A}(\omega) \mathcal{A}^2(\omega^{\prime})
	\nonumber \\
	&\hspace{-10mm} 
	\times \left( \Gamma_L \frac{\partial f_{L}(\omega^{\prime})}{\partial \mu_L} 
	\frac{\partial f_{R}(\omega)}{\partial \mu_R} + \Gamma_R \frac{\partial f_{L}(\omega)}{\partial \mu_L} 
	\frac{\partial f_{R}(\omega^{\prime})}{\partial \mu_R} \right). \label{eqn:integral_kernel_V}
\end{align}
The kernel $\Pi_V(\mu_L,\mu_R)$ has the same form as $\Pi_T(T_L,T_R)$ (see Eq.~(\ref{eqn:integral_kernel_T})) except that the temperature derivative is replaced with the electrochemical-potential derivative.

Although other kinds of the parameter driving (e.g. $T_L$-$\mu_L$ driving) are possible, we focus on the two simple cases given above, and clarify difference between them in the next section.
We note that once the integral kernels are defined as Eqs.~(\ref{eqn:integral_kernel_T}) and (\ref{eqn:integral_kernel_V}), then the anticlockwise trajectory must be taken in the parameter space of $(T_L,T_R)$ or $(\mu_L,\mu_R)$.
If the trajectory is set to clockwise, the sign of the integral kernels becomes opposite.

\section{\label{sec:Numerical_calculation}Numerical Result}
\label{sec:numerical}

In this section, we show the contour plots of the integral kernel of both of the temperature-driven pumping and the electrochemical-potential-driven pumping, and discuss their qualitative properties.
We also estimate the pumping current averaged in one cycle for both types of pumping, and show that it is large enough for experimental observation.

\subsection{Temperature-driven pumping}

\begin{figure}
\begin{center}
\includegraphics[clip,width=8.0cm]{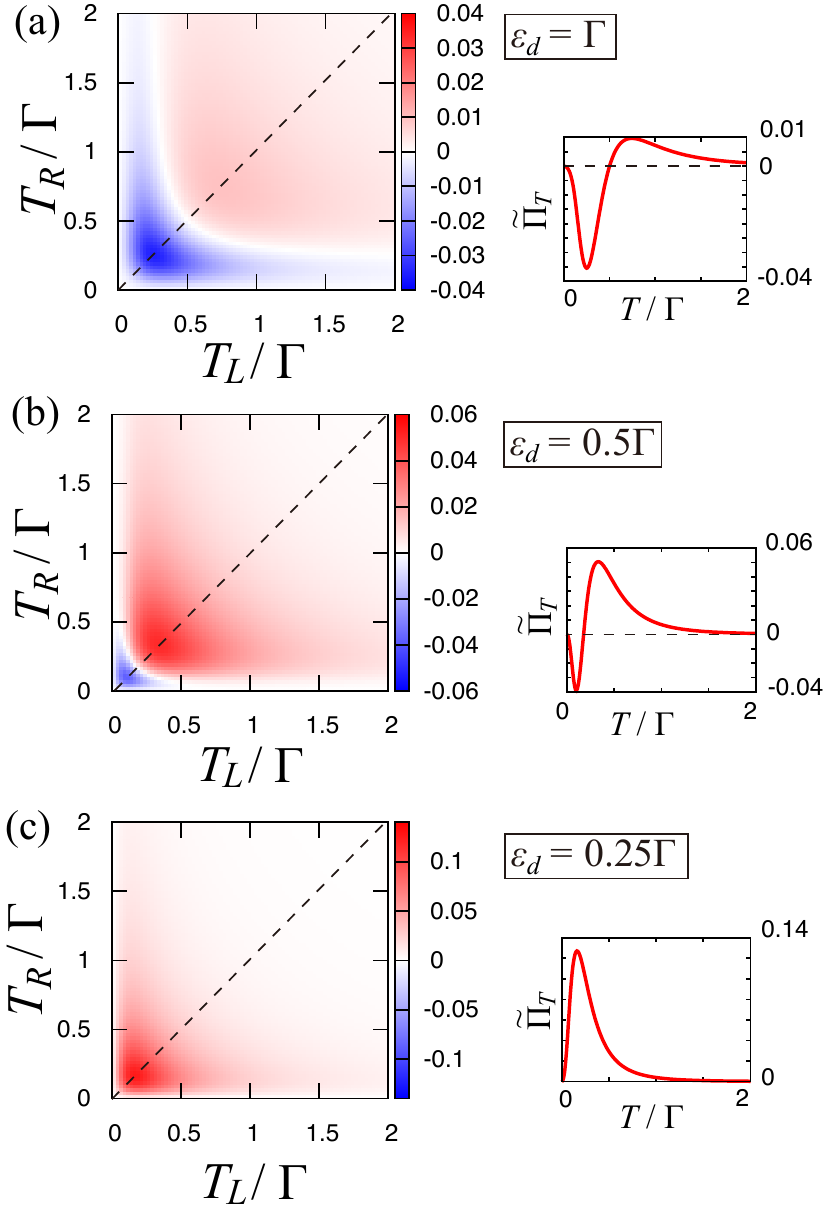}
\end{center}
\caption{\label{fig:cpdata} 
Left panels: Contour plots of the normalized integral kernel $\tilde{\Pi}_T(T_L,T_R) = (\Gamma^3/U) \Pi_T (T_L, T_R)$ for (a) $\epsilon_d = \Gamma$, (b) $\epsilon_d = 0.5\Gamma$, and (c) $\epsilon_d = 0.25\Gamma$. 
The symmetric coupling ($\Gamma_L = \Gamma_R = \Gamma/2$) is assumed. 
As the QD energy level $\epsilon_d$ decreases, the area in which the kernel takes positive values shrinks, and disappears for $\epsilon_d < \Gamma/(2\sqrt{3}) \simeq 0.29\Gamma$.
This indicates that one-way pumping is realized for $\epsilon_d < \Gamma/(2\sqrt{3})$.
Right panels: Plots of the normalized integral kernel on the line $T_L = T_R \equiv T$.
}
\end{figure}

First, we consider the temperature-driven pumping by setting $\mu_L(t) = \mu_R(t) = 0$.
When the sign of $\epsilon_d$ is reversed, the integral kernel changes only its sign keeping its magnitude.
Therefore, it is sufficient to consider the integral kernel only for $\epsilon_d > 0$.
The left panels of Fig.~\ref{fig:cpdata} show the contour plots of the normalized integral kernel defined by
\begin{align}
\tilde{\Pi}_T (T_L, T_R) \equiv \frac{\Gamma^3}{U} \Pi_T(T_L,T_R) ,
\end{align}
for the symmetric coupling ($\Gamma_L = \Gamma_R = \Gamma/2$) with three different positive QD energy levels, $\epsilon_d = \Gamma$, $0.5 \Gamma$, and $0.25 \Gamma$. 
These contour plots indicate the strength of charge pumping; as explained in Sect.~\ref{sec:kernelrep}, arbitrary adiabatic temperature modulation is expressed by a trajectory in the $(T_L, T_R)$ space, and the pumped charge for one cycle is a surface integral of the integral kernel inside the trajectory. 
When the QD energy level is far from the Fermi level ($\epsilon_d = \Gamma, 0.5\Gamma$), the integral kernel becomes positive (negative) in the high-temperature (low-temperature) side.
As $\epsilon_d$ decreases, the region of the negative kernel becomes narrower, and disappears when $\epsilon_d$ is below a threshold value, which is shown to be $\Gamma/(2\sqrt{3}) \simeq 0.29\Gamma$ later.
Actually, the integral kernel is always positive for $\epsilon_d = 0.25 \Gamma$ as shown in the left panel in Fig.~\ref{fig:cpdata}~(c).
The right panels of Fig.~\ref{fig:cpdata} shows the integral kernel on $T_L = T_R \equiv T$.
The integral kernel has one positive peak and one negative dip for $\epsilon_d = \Gamma, 0.5\Gamma$, while it has only one positive peak for $\epsilon_d = 0.25\Gamma$. 

\begin{figure}
\includegraphics[clip,width=7.0cm]{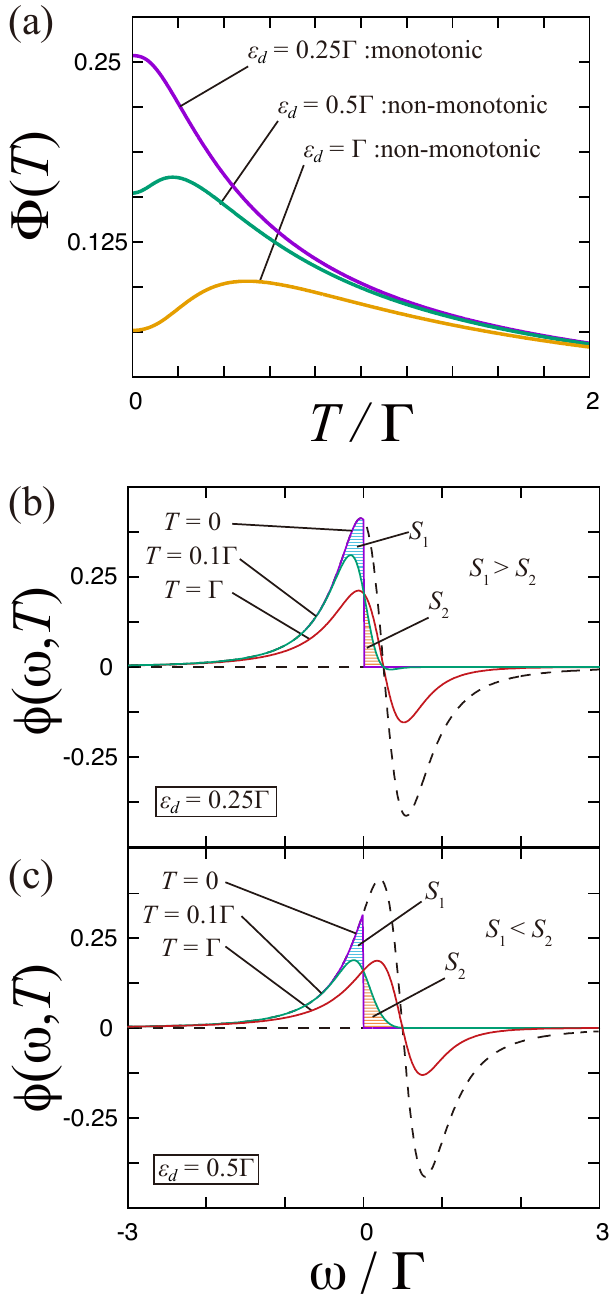}
\caption{\label{fig:scdp} (a) Temperature-dependence of $\Phi(T)$ for three QD energy levels, $\epsilon_d=0.25\Gamma$, $0.5\Gamma$ and $\Gamma$. 
For $\epsilon_d=0.25\Gamma$, $\Phi(T)$ is monotonic decreasing function while it is non-monotonic function for $\epsilon_d=0.5\Gamma, \Gamma$.
(b) Plots of the integrand $\phi(\omega,T)$ as a function of $\omega$ for $T=0$, $0.1\Gamma$ and $\Gamma$ for $\epsilon_d = 0.25 \Gamma$.
An integral of $\phi(\omega,T)$ in the range of $\omega \in [-\infty:\infty]$ gives $\Phi(T)$ except for the constant coefficient (see Eq.~(\ref{eq:defintegrand})).
The dashed line shows $\partial_{\omega} \mathcal{A}(\omega;\epsilon_d)$.
(c) The same plots for $\epsilon_d = 0.5 \Gamma$.
The difference, $\Phi(T=0.1\Gamma)-\Phi(\omega,T=0)$, is given by $S_2-S_1$, where $S_1 = |\int_{-\infty}^0 d\omega (\phi(\omega,T=0.1\Gamma)-\phi(\omega,0)) |$ is negative contribution of holes for $\omega < 0$ and $S_2 = \int_0^\infty d\omega (\phi(\omega,T=0.1\Gamma)-\phi(\omega,0))$ is positive contribution of electrons for $\omega > 0$.
At $T=\Gamma$, the Fermi distribution function is sufficiently broadened and the negative part of $ \partial_{\omega} \mathcal{A}(\omega;\epsilon_d)$ starts to contribute.
Thus, $\Phi(T)$ approaches to zero at high temperatures.
}
\end{figure}

In order to understand this qualitative change across $\epsilon_d = \Gamma/(2\sqrt{3})$, we rewrite the integral kernel given in Eq.~(\ref{eqn:integral_kernel_T}) as
\begin{align}
\tilde{\Pi}_T(T_L,T_R) &= -e\Gamma^2 \biggl(\partial_{T_L} \Phi(T_L) \partial_{T_R} \Psi(T_L,T_R) \nonumber \\
& \hspace{5mm}+ \partial_{T_R} \Phi(T_R) \partial_{T_L} \Psi(T_L,T_R)\biggr), 
\label{eq:KernelIntegral} \\
\Phi(T) &= \frac{2\Gamma_L \Gamma_R}{\Gamma} 
\int d\omega \partial_{\omega} \mathcal{A}(\omega;\epsilon_d) f(\omega; T), \\
\Psi(T_L, T_R) &= \Gamma \int d\omega \pi \mathcal{A}^2(\omega;\epsilon_d) f_{\rm eff}(\omega; T_L, T_R).
\end{align}
Here, the Fermi distribution function is denoted with $f(\omega;T)$ to stress the temperature dependence, $\Phi(T)$ is an integral related to the steady current as
\begin{align}\label{eqn:pc_two_part}
	\partial_{\epsilon_d} J_L^{\mathrm{steady}}(\epsilon_d,T_L,T_R) = e(\Phi(T_L) - \Phi(T_R)),
\end{align}
and $\Psi(T_L, T_R)$ is an integral related to the change of the occupation number due to the time delay, $\langle n_s(t) \rangle_0^{(1)}$, as
\begin{align}
\langle n_s(t) \rangle_0^{(1)} &= -\frac{1}{\Gamma} \frac{d\Psi(T_L(t),T_R(t))}{dt} \nonumber \\
&= - \frac{\partial \Psi(T_L(t),T_R(t))}{\partial T_L} \frac{\dot{T}_L(t)}{\Gamma}- \frac{\partial \Psi(T_L(t),T_R(t))}{\partial T_R} \frac{\dot{T}_R(t)}{\Gamma}.
\end{align}
As given in Eq.~(\ref{eq:KernelIntegral}), the sign of the integral kernel is determined by the temperature derivatives of $\Phi(T)$ and $\Psi(T_L, T_R)$.
We note that $\Psi(T_L, T_R)$ is a monotonically increasing function of both the temperatures, $T_L$ and $T_R$, i.e., satisfies
\begin{align}
\partial_{T_L} \Psi(T_L, T_R) > 0, \quad \partial_{T_R} \Psi(T_L, T_R) > 0,
\label{eq:inequilityPsi}
\end{align}
for $\epsilon_d > 0$.
As a result, the sign change of the integral kernel is caused only by that of $\partial_T \Phi(T)$.
In Fig.~\ref{fig:scdp}~(a), we show a plot of $\Phi(T)$ as a function of temperature for $\epsilon_d=0.25\Gamma$, $0.5\Gamma$ and $\Gamma$. 
As seen in the figure, $\Phi(T)$ is a non-monotonic function of $T$ for $\epsilon_d=\Gamma, 0.5\Gamma$, and the sign of $\partial_T \Phi(T)$ changes at a certain temperature.
On the other hand, $\Phi(T)$ is a monotonically decreasing function of $T$ for $\epsilon_d=0.25\Gamma$, and the sign of $\partial_{T} \Phi(T)$ is always negative.

The non-monotonic behavior of $\Phi(T)$ causes the sign change of the integral kernel as follows.
When $\Phi(T)$ is a monotonically decreasing function of $T$, both $\partial_{T_L} \Phi(T_L)$ and $\partial_{T_R} \Phi(T_R)$ are negative for arbitrary value of $T_L$ and $T_R$.
By combining this with Eq.~(\ref{eq:inequilityPsi}), we can conclude that the integral kernel is always positive from Eq.~(\ref{eq:KernelIntegral}).
When $\Phi(T)$ is a non-monotonic function of $T$, $\Phi(T)$ has a maximum value at a certain temperature $T=T_{\rm max}$.
If we set both of $T_L$ and $T_R$ as larger (smaller) than $T_{\rm max}$, the integral kernel becomes positive (negative) from Eq.~(\ref{eq:KernelIntegral}) since both of $\partial_{T_L} \Phi(T_L)$ and $\partial_{T_R} \Phi(T_R)$ becomes negative (positive).
Therefore, we can conclude that the integral kernel must change its sign at least once.
Thus, the non-monotonicity of $\Phi(T)$ is the origin of the sign change of the integral kernel.
When $T_L \le T_{\rm max} \le T_R$ or $T_R \le T_{\rm max} \le T_L$, the integral kernel can be either positive or negative, and determination of its sign needs detailed analysis of Eq.~(\ref{eq:KernelIntegral}).
We stress that since $\Phi(T)$ is related to $\partial_{\epsilon_d} J_L^{\mathrm{steady}}(\epsilon_d,T_L,T_R)$ as given in Eq.~(\ref{eqn:pc_two_part}), its behavior reflects the response of the steady current with respect to change of the QD energy level.

To discuss the origin of the non-monotonic (monotonic) behavior of $\Phi(T)$ for large (small) values of $\epsilon_d$, we define the integrand $\phi(\omega,T)$ as
\begin{align}
\Phi(T) &= \frac{2\Gamma_L \Gamma_R}{\Gamma^3}  \int d\omega \phi(\omega,T), \label{eq:defintegrand} \\
\phi(\omega,T) &= \Gamma^2 \partial_{\omega} \mathcal{A}(\omega;\epsilon_d) f(\omega; T),
\end{align}
and show the $\omega$-dependence of $\phi(\omega,T)$ for three different temperatures, $T=0$, $0.1\Gamma$, and $\Gamma$ in Fig.~\ref{fig:scdp}~(b) and (c).
In the figures, we also show $\Gamma^2 \partial_\omega \mathcal{A}(\omega;\epsilon_d)$ as dashed line, which has a peak at $\omega = \epsilon_d -\Gamma/(2\sqrt{3})$.
There are two cases.
In the case of $0 < \epsilon_d < \Gamma/(2\sqrt{3})$, the Fermi energy ($\omega = 0$) is located above the frequency of the peak of $\partial_\omega \mathcal{A}(\omega)$ as shown in Fig.~\ref{fig:scdp}~(b).
At zero temperature, $\phi(\omega,T)$ is a product of $\partial_{\omega} \mathcal{A}(\omega;\epsilon_d)$ and the step function $\Theta(-\omega)$, and therefore, $\Phi(T=0)$ is proportional to an integral of $\partial_{\omega} \mathcal{A}(\omega;\epsilon_d)$ in the range of $\omega \in [-\infty, 0]$.
At low temperatures (e.g. at $T=0.1\Gamma$ shown in the figure), the jump of the distribution function is smeared by thermal broadening.
The change of the integral, $\Phi(T)-\Phi(0)$, is then given by a sum of negative contribution by holes at $\omega < 0$  (denoted with $S_1$ in the figure) and positive contribution by electrons at $\omega >0$ (denoted with $S_2$ in the figure).
Since $\partial_{\omega} \mathcal{A}(\omega;\epsilon_d)$ is larger in the hole side than in the electron side ($S_1 > S_2$), $\Phi(T)$ decreases as the temperature rises from zero.
At higher temperatures (e.g. at $T=\Gamma$ shown in the figure), $\Phi(T)$ decreases further because the negative part of $\partial_{\omega} A(\omega;\epsilon_d)$ at high energies ($\omega > \epsilon_d$) starts to contribute.
Thus, in this case, $\Phi(T)$ monotonically decreases as the temperature increases.
In the case of $\Gamma/(2\sqrt{3}) < \epsilon_d$, the Fermi energy ($\omega = 0$) is located below the peak frequency of $\partial_\omega \mathcal{A}(\omega)$ as shown in Fig.~\ref{fig:scdp}~(c).
In this case, positive contribution by electrons exceeds negative contribution by holes ($S_1 < S_2$) at low temperatures since $\partial_{\omega} \mathcal{A}(\omega;\epsilon_d)$ is larger in the electron side than in the hole side.
Then, the change of the integral, $\Phi(T)-\Phi(0)$, becomes positive at low temperature.
This indicates that $\Phi(T)$ increases as the temperature rises from zero.
However, at higher temperatures, $\Phi(T)$ is a decreasing function of $T$ because the negative part of $\partial_{\omega} A(\omega;\epsilon_d)$ at high energies ($\omega > \epsilon_d$) starts to contribute.
Thus, in this case, $\Phi(T)$ is a non-monotonic function of $T$.
We stress that this qualitative change of the behavior of $\Phi(T)$ clearly originates from the level broadening due to strong coupling to the reservoirs, i.e., the finite line width of the spectrum function $\mathcal{A}(\omega)$.

In summary, we have related the sign change of the integral kernel to the non-monotonicity of $\Phi(T)$.
We point out that in the limit of the weak dot-reservoir coupling ($\Gamma \ll T, \epsilon_d$), $\Phi(T)$ is proved to be non-monotonic, and the sign change of the kernel always occurs.
Therefore, the disappearance of the sign change of the kernel observed for small $\epsilon_d$ is a direct consequence of level broadening effect due to strong dot-reservoir coupling.
This is a new feature of the present pumping scheme, which has not been studied in the previous theoretical works on a basis of weak lead-dot coupling.

\subsection{Electrochemical-potential-driven pumping}

\begin{figure}
\includegraphics[clip,width=9.0cm]{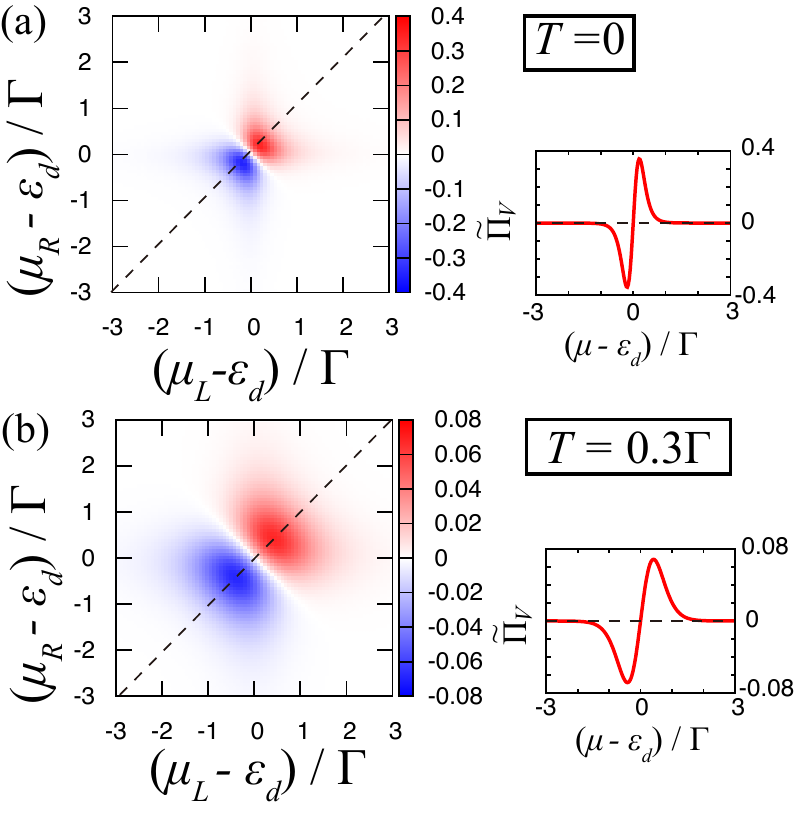}
\caption{\label{fig:cpdata_mu} 
Left panels: Contour plots of the normalized integral kernel $\tilde{\Pi}_V(\mu_L, \mu_R) = (\Gamma^3/U) \Pi_V (\mu_L, \mu_R)$ for (a) $T_L = T_R=0$ and (b) $T_L = T_R=0.3\Gamma$.
The symmetric coupling ($\Gamma_L = \Gamma_R = \Gamma/2$) is assumed.
At both temperatures, the integral kernel has symmetries, $\Pi_V(\mu_L-\epsilon_d,\mu_R-\epsilon_d) = \Pi_V(\mu_R-\epsilon_d, \mu_L-\epsilon_d) = - \Pi_V(-(\mu_L-\epsilon_d),-(\mu_R-\epsilon_d))$, and changes its sign on the line $\mu_L -  \epsilon_d = -(\mu_R - \epsilon_d)$.
The integral kernel takes a maximum value and a minimum value on the line $\mu_L = \mu_R$.
Right panels: Plots of the normalized integral kernel on the line $\mu_L = \mu_R$.
}
\end{figure}

Next, we show contour plots of the integral kernel for electrochemical-potential-driven pumping for the symmetric coupling ($\Gamma_L = \Gamma_R = \Gamma/2$).
We define the normalized integral kernel by
\begin{align}
\tilde{\Pi}_V (\mu_L, \mu_R) \equiv \frac{\Gamma^3}{U} \Pi_V(\mu_L,\mu_R),
\end{align}
and show its contour plots in the left panes of Fig.~\ref{fig:cpdata_mu} for two different sets of the reservoir temperatures, (a) $T_L = T_R = 0$ and (b) $T_L = T_R = 0.3 \Gamma$.
As seen in the contour plots, the integral kernel has a symmetry relation $\Pi_V(\mu_L,\mu_R) = \Pi_V(\mu_R,\mu_L)$, and takes a maximum value and a minimum value on the line $\mu_L=\mu_R$.
The normalized integral kernel on the line of $\mu_L = \mu_R$ is shown in the right panels of Fig.~\ref{fig:cpdata_mu}.
At zero temperature, the kernel takes a maximum value at $\mu_L-\epsilon_d=\mu_R-\epsilon_d= \Gamma/(2\sqrt{7})$ ($\simeq 0.19\Gamma$), and takes a minimum value at $\mu_L-\epsilon_d=\mu_R-\epsilon_d= -\Gamma/(2\sqrt{7})$.
At finite temperatures, due to thermal broadening, the positions of the peak and the dip move away from the origin, and their width is broadened.

As indicated in the left panes of Fig.~\ref{fig:cpdata_mu}, the integral kernel always changes its sign on the line $\mu_L - \epsilon_d = -(\mu_R -\epsilon_d)$.
This property is proved as follows. 
In the same manner as temperature-driven pumping, the integral kernel is rewritten as
\begin{align}
	\tilde{\Pi}_V(\mu_L,\mu_R) &= -e\Gamma^2 \biggl(\partial_{\mu_L} \Phi(\mu_L) \partial_{\mu_R} \Psi(\mu_L,\mu_R) \nonumber \\
	& \hspace{5mm}+ \partial_{\mu_R} \Phi(\mu_R) \partial_{\mu_L} \Psi(\mu_L,\mu_R)\biggr),  \label{eq:PiVsym1} \\
	\Phi(\mu) &= \frac{2\Gamma_L \Gamma_R}{\Gamma} \int d\omega \partial_{\omega} \mathcal{A}(\omega;0) f(\omega; \mu-\epsilon_d), \label{eq:PiVsym2}\\
	\Psi(\mu_L, \mu_R) &= \Gamma \int d\omega \pi \mathcal{A}^2(\omega;0) f_{\rm eff}(\omega; \mu_L-\epsilon_d, \mu_R-\epsilon_d).
	\label{eq:PiVsym3}
\end{align}
Here, the Fermi distribution function and the effective distribution function are denoted with $f(\omega;\mu)$ and $f_{\rm eff}(\omega;\mu_L,\mu_R)$, respectively, to stress the electrochemical-potential dependence, and the integral variable $\omega$ is shifted by $\epsilon_d$.
From Eqs.~(\ref{eq:PiVsym1})-(\ref{eq:PiVsym3}), it is straightforward to prove the symmetry relation $\Pi_V(\mu_L,\mu_R) = \Pi_V(\mu_R,\mu_L)$.
Furthermore, by using the relations
\begin{align}
f(\omega; \mu-\epsilon_d) &=  1 - f(-\omega;\epsilon_d-\mu), \\
f_{\rm eff}(\omega;\mu_L-\epsilon_d,\mu_R-\epsilon_d) &= 1 - f_{\rm eff}(-\omega;\epsilon_d-\mu_L,\epsilon_d-\mu_R), 
\end{align}
we can also prove $\Pi_V(\mu_L,\mu_R) = -\Pi_V(2\epsilon_d-\mu_L,2\epsilon_d-\mu_R)$.
From these symmetry relations, we can conclude that the sign of the integral kernel changes on the line $\mu_L - \epsilon_d = -(\mu_R -\epsilon_d)$.

\subsection{Estimate of the averaged pumping current}
\label{sec:estimate}

Let us estimate the pumping current averaged in one cycle using a set of the realistic model parameters.
Before the estimate, we give some remarks on the range of the model parameters, to which the present formulation can be applied.
Since the pumped charge is evaluated up to the first order of Coulomb interaction, the present result is applicable only for a small value of $U$.
For the Anderson model, the first-order perturbation is known to be sufficiently good when $U \ll \pi \Gamma$.
In the following estimate, we use $U=\Gamma$, which is small enough to give a quantitatively correct result (see Ref.~\citen{Zlatic89} for example).
In addition, we have employed the adiabatic approximation justified for the low-frequency pumping $\mathcal{T}^{-1} \ll \Gamma$; if the pumping frequency $\mathcal{T}^{-1}$ is larger than $\Gamma$, the non-adiabatic effect starts to contribute.
For the purposes of rough estimation, we set $\mathcal{T}^{-1} = 1\mathrm{GHz}$.

\begin{figure}
\includegraphics[clip,width=7.0cm]{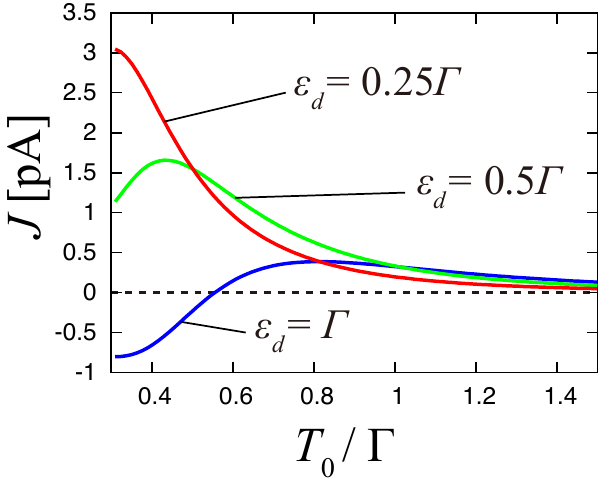}
\caption{\label{fig:cidatas} Estimated pumping current averaged in one period of the temperature driving 
$(T_L , T_R) = (T_0 +  T_1 \cos (2\pi \mathcal{T} t), T_0 + T_1 \sin(2\pi \mathcal{T} t))$ for the symmetric coupling 
($\Gamma_L =\Gamma_R=\Gamma/2$).
The three curves show the pumping current as a function of $T_0$ for three different values of $\epsilon_d$.
The other parameters are set to $U=\Gamma$, $T_1 = 0.3\Gamma$ and $\mathcal{T}^{-1} =1 \mathrm{GHz}$.
The average pumping current is always positive for $\epsilon_d = 0.25 \Gamma$ while it becomes negative at low temperatures for $\epsilon_d = \Gamma$ in consistent with the contour plots of the integral kernel shown in the right panels of Fig.\ref{fig:cpdata}.
Although the sign of the integral kernel becomes negative at low temperatures for $\epsilon_d = 0.5 \Gamma$, the average pumping current is alway positive because the amplitude of the temperature ($T_1 = 0.3\Gamma$) is too large to see the sign change at low temperatures.
}
\end{figure}

First, we estimate the pumping current averaged in one period, $J = \delta Q_{L,1}^{\mathrm{pump}}/\mathcal{T}$, induced by the temperature driving by setting $\mu_L = \mu_R = 0$ and
\begin{align}
	(T_L , T_R) 
	= (T_0 + T_1 \cos (2\pi t/\mathcal{T}), T_0 + T_1\sin (2\pi t/\mathcal{T})). \label{eq:TempModulation} 
\end{align}
Fig.~\ref{fig:cidatas} shows the average pumping current for three values of $\epsilon_d$ as a function of $T_0$ for the symmetric coupling  ($\Gamma_L =\Gamma_R = \Gamma/2$) and $T_1 = 0.3\Gamma$.
The pumping current is of the order of $1\ {\rm pA}$, which are measurable in a standard experimental setup.
In the present calculation, the pumping current is proportional to the pumping frequency as far as the condition for the adiabatic approximation holds well ($\mathcal{T}^{-1} \ll \Gamma$).
Therefore, the pumping frequency can be set to smaller value if one allows a smaller pumping current for detection.
The estimated current in Fig.~\ref{fig:cidatas} also reflects the qualitative property of the integral kernel shown in Fig.~\ref{fig:cpdata};
it is always positive for $\epsilon_d = 0.25 \Gamma$ and changes its sign for $\epsilon_d = \Gamma$.
We note that for $\epsilon_d = 0.5 \Gamma$ the sign change of the pumping current is not observed in Fig.~\ref{fig:cidatas}.
This is because temperature modulation amplitude is too large to pick up the sign change, and the sign change can be observed when the amplitude is sufficiently small.

\begin{figure}
\includegraphics[clip,width=7.0cm]{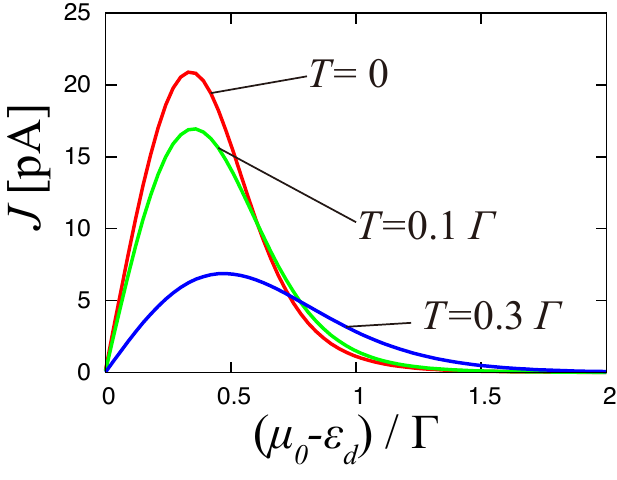}
\caption{\label{fig:cidatas_mu} The estimated average pumping current under the electrochemical potential modulation $(\mu_L , \mu_R) = (\mu_0 +  \mu_1 \cos(2\pi \mathcal{T} t), \mu_0 + \mu_1 \sin(2\pi \mathcal{T} t))$ for the symmetric coupling ($\Gamma_L =\Gamma_R=\Gamma/2$).
The three curves show the pumping current as a function of $\mu_0$ for three different temperatures of $T$.
The other parameters are set to $U=\Gamma$, $\mu_1 = 0.5\Gamma$ and $\mathcal{T}^{-1} =1 \mathrm{GHz}$.
As temperature increases, the peak of the pumping current is suppressed by thermal broadening, and the peak position shifts toward a higher chemical potential.
}
\end{figure}

Next, we estimate the average pumping current induced by the electrochemical-potential driving by setting $T_L = T_R = T$ and
\begin{align}
	(\mu_L , \mu_R)
	= (\mu_0 + \mu_1 \cos (2\pi t/\mathcal{T}), \mu_0 + \mu_1\sin (2\pi t/\mathcal{T})). 
	\label{eq:ChemModulation}
\end{align}
Fig.~\ref{fig:cidatas_mu} shows the average pumping current at three values of $T$ as a function of $\mu_0-\epsilon_d$ for the symmetric coupling ($\Gamma_L =\Gamma_R = \Gamma/2$) and $\mu_1=0.5\Gamma$.
Note that the average current $J$ is an antisymmetric function of $\mu_0-\epsilon_d$ (see also the contour plot shown in Fig.~\ref{fig:cpdata_mu}).
As seen in Fig.~\ref{fig:cidatas_mu}, the pumping current is of the order of $10\ {\rm pA}$, which are measurable in a standard experimental setup.

\section{\label{sec:discussion}Discussion}

In this section, we discuss the experimental realization of the present result. 
We also discuss the relation between the time delay and the AC response of the QD.

\subsection{Experimental realization}

Although quick control of the reserver electrochemical potential during charge pumping is expected to be realized rather easily in experiments, quick control  of the reservoir temperatures may be an experimental challenge.
There are several methods to achieve a fast control of the reservoir temperatures.
One possible strategy is to introduce a periodic series of microwave pulses (or a periodic amplitude modulation of a microwave) into a circuit to warm up the electrons in the reservoirs.
If the thermalization time in the reservoirs is much smaller than the pumping period $\mathcal{T}$, the situation treated in our calculation will be realized.
Another possible strategy is to use a non-equilibrium energy distribution of electrons in a channel near the point contact~\cite{Altimiras10}. 
As the non-equilibrium energy distribution generated at the point contact relaxes into a thermal distribution with an effective temperature after thermalization.
Since the non-equilibrium energy distribution is controlled by the transmission probability of the point contact, the effective temperature can be quickly tweaked using the gate voltage at the point contact~\cite{footnote:discussion}.

\subsection{Relation to the AC response}
\label{sec:ACresponse}

The physical meaning of the time delay $\delta t$ can be discussed by relating it to the dynamic response of the QD under AC modulation of the temperature and the electrochemical potential.
The linear AC admittance of a QD system coupled with one reservoir is effectively described by a parallel circuit with a capacitance and a resistance, operating within a low-frequency limit~\cite{Buttiker93b,Nigg06,Gabelli06,Gabelli07}.
These circuit elements are known as the dynamical capacitance and the dynamical resistance, respectively.
Recently, the concept of dynamical circuit elements has been extended to coherent transport under AC temperature modulation~\cite{Lim13}.
It is possible to express the time delay $\delta t$ by a linear combination of time constants determined by these circuit elements (see Appendix\ref{App:DynamicalResistance}).
Accordingly, the time delay can be regarded as a characteristic relaxation time determined by the effective circuit elements describing the AC transport in a QD in the coherent regime.

\section{\label{sec:summary}Summary} 

We have studied adiabatic charge pumping driven by temperatures and electrochemical potentials of reservoirs via a single-level QD strongly coupled to two reservoirs.
We have derived a formula of adiabatic charge pumping, and have shown that charge pumping occurs in the presence of Coulomb interaction within its first-order perturbation.
In the present calculation, the effective QD energy level with the Hartree-type correction responses to the time-dependent reservoir parameters via the change of the occupation of the QD.
We have shown that a time delay exists in this response, and that this delay causes rectification of the charge current, resulting in net charge pumping.
We have also rewritten the formulas of pumped charge into an area integral of an integral kernel for temperature-driven pumping and electrochemical-potential-driven pumping, respectively, and have discussed their properties in detail.
Especially in temperature-driven pumping, we have shown that one-way pumping can be realized when the QD energy level is near the Fermi level while the sign of the pumping changes depending on temperatures when the QD energy level is far from the Fermi level.
This one-way pumping originates from the broadening effect of the QD, and is a new feature obtained only in the QD strongly coupled to the reservoirs.
Finally, we have estimated pumping current by choosing realistic experimental parameters, and have shown that it is on the order of pA, which is measurable in experiment.

In this paper, we have studied charge pumping in a simple setup with a single-level QD up to the leading order of Coulomb interaction.
It is a future problem to study a phase-sensitive setup such as the Aharonov-Bohm effect and the Fano effect to examine how a coherent nature of electron transport affects the present adiabatic pumping.
It is another future problem to formulate adiabatic pumping in the presence of a strong Coulomb interaction. 
With such formalism, we would be able to investigate many body effects characteristic for coherent transport (the Kondo effect, for example) on adiabatic pumping by temperature and electrochemical potential modulation.
Also, we would be able to extend the present formalism to energy pumping.
It is an interesting challenge for the nonequillibrium thermodynamics in mesoscopic systems to discuss heat transport and define nonequillibrium entropy of the mesoscopic systems.

\section*{Acknowledgments}
The authors are grateful to T. Yokoyama for critically reading the manuscript and suggesting substantial improvements. The authors also thank R. Sakano, G. Tatara, C. Uchiyama, T. Martin, A. Shitade, T. Arakawa and K. Kobayashi for helpful discussions.
This work was supported by Japan Society for the Promotion of Science KAKENHI Grants No. JP24540316 and No. JP26220711.
M.H. acknowledges financial support provided by the Advanced Leading Graduate Course for Photon Science.

\appendix

\section{\label{apx:selfenergy} Calculation of the Self-Energy}

In this appendix, we derive the expressions for the self-energy given in
Eqs.~(\ref{eq:TempModulatedSelfEnergy1})-(\ref{eq:TempModulatedSelfEnergy2}).
The GFs for electrons with wavenumber $k$ and spin $s$ in the isolated reservoir $r$ are given by
\begin{align}
        g_{r ks}^R(t,t^{\prime}) &= - i \Theta (t-t^{\prime}) e^{-i \epsilon_{r ks}(t -t^{\prime})} = \left( g_{r ks}^A(t^{\prime},t) \right)^{\ast}, \\
        g_{r ks}^<(t,t^{\prime}) &= i\int d\omega f(\omega) \delta( \omega - \epsilon_{r ks} ) e^{-i \epsilon_{r ks}(t -t^{\prime})},
\end{align}
where $f(\omega)= [e^{\omega/T_{\mathrm{ref}}} + 1]^{-1}$.
Taking the wide-band limit, the retarded self-energy is calculated from Eq.~(\ref{eq:self1}) as
\begin{align}
        \Sigma_s^R(t_1,t_2) &= -i \Theta(t_1-t_2) \sum_{r} \Gamma_{r}
        \prod_{i=1}^{2} \sqrt{1+ \dot{B}_{r}(t_i)} \nonumber \\
        & \hspace{5mm} \times \delta( t_1 - t_2 + B_{r}(t_1) - B_{r}(t_2) ) e^{-i(\Lambda_r(t_1) - \Lambda_r(t_2))}
        \nonumber \\ \label{eqn:self_ene_r}
        &= -i \sum_{r} \frac{\Gamma_{r}}{2} \delta(t_1 -t_2).
\end{align}
In the last line, we have used the fact that $t_1 - t_2 + B_{r}(t_1) - B_{r}(t_2) = 0$ has only one solution $t_1 = t_2$ under the condition (\ref{eq:positiveness}). 
We note that the time-dependence of the dot-reservoir coupling does not change
the retarded self-energy with arbitrary time dependence of external fields for the wide-band limit.

The lesser component of the self-energy is calculated from Eq.~(\ref{eq:self2}) in a similar way:
\begin{align}
        \Sigma_s^<(t_1,t_2) &= i  \sum_{r} \frac{\Gamma_{r}}{2\pi}
        \prod_{i=1}^{2} \sqrt{1+ \dot{B}_{r}(t_i)} \nonumber \\
        &  \times \int d\omega f(\omega) e^{-i\omega ( t_1 - t_2 + B_{r}(t_1) - B_{r}(t_2) )} \nonumber \\
        &  \times e^{-i(\Lambda_r(t_1) - \Lambda_r(t_2))}.
\end{align}
Supposing that $B_r(t) = \tilde{B}_r t$ and $\Lambda_r(t) = \tilde{\mu}_r t$, the self-energy becomes
\begin{align}
        \Sigma_s^<(t_1,t_2) &= i  \sum_{r} \frac{\Gamma_{r}}{2\pi} (1+ \tilde{B}_{r}) \nonumber \\
        & \qquad \times \int d\omega f(\omega) e^{-i[\omega(1+\tilde{B}_r) + \tilde{\Lambda}_r]( t_1 - t_2) } \nonumber \\
        & = i \sum_{r} \frac{\Gamma_{r}}{2\pi} \int d\omega f \left( \frac{\omega - \tilde{\mu}_r}{1+\tilde{B}_r} \right)
        e^{-i\omega ( t_1 - t_2) }.
\end{align}

\section{\label{apx:rep_of_Q0Q1}An explicit form of $\delta Q_{r,0}$ and $\delta Q_{r,U}$}

The zeroth-order (noninteracting) contribution $\delta Q_{r,0}$ is given by
\begin{align}
	\delta Q_{r,0} &= \delta Q_{r,0,{\rm A}} + \delta Q_{r,0, {\rm B}}, 
	\label{eqn:noninteracting_tot} \\
	\delta Q_{r,0,{\rm A}} &=-4e \int_{t_i}^{t_f} dt \int dt_1\int \frac{d\omega}{2\pi} \Gamma_r \nonumber \\
	&\hspace{-10mm} \times \mathrm{Re} \Bigl[ i G_{0}^R(t,t_1) f(\omega) e^{-i\omega(t_1-t)} C_{r}(\omega,t_1,t) \Bigr], 		\label{eqn:noninteracting_A} \\
	\delta Q_{r,0,{\rm B}} &= 2e \sum_{r^{\prime}} \int_{t_i}^{t_f} dt \int dt_1dt_2 \int \frac{d\omega}{2\pi} 
	\Gamma_r \Gamma_{r^{\prime}} \nonumber \\
	&\hspace{-10mm} \times \mathrm{Re} \Bigl[ G_{0}^R(t,t_1) G_{0}^A(t_2,t) f(\omega) e^{-i\omega(t_1-t_2)} C_{r^{\prime}}(\omega,t_1,t_2) \Bigr],
	\label{eqn:noninteracting_B}
\end{align}
where $\delta Q_{r,0,A}$ and $\delta Q_{r,0,B}$ correspond to the first and the second term in the bracket
of Eq.~(\ref{eqn:acm_cur}), respectively. In the same way, the first-order (interacting) contribution 
$\delta Q_{r,1}$ is given by
\begin{align}
	\delta Q_{r,U} &= \delta Q_{r,U,{\rm A}} + \delta Q_{r,U,{\rm B}}, \\
	\delta Q_{r,U,{\rm A}} &=   -4eU \sum_{r^{\prime}} \int \frac{d\omega_1 d\omega_2}{(2\pi)^2} \int_{t_i}^{t_f} dt \int dt_1 \cdots dt_4  \Gamma_{r^{\prime}} \Gamma_{r} \nonumber \\
	&\hspace{-10mm}\times \mathrm{Re} \Bigl[ i \mathcal{G}_A(t_1,t_2,t_3,t_4) f(\omega_1) f(\omega_2) e^{-i\omega_1 t_1}e^{-i\omega_2(t_3-t_4)}   \nonumber \\
	& \hspace{-10mm}\qquad \qquad \times C_{r}(\omega_1,t_1+t,t) C_{r^{\prime}}(\omega_2,t_3+t,t_4+t) \Bigr],
	\label{eqn:deltaQ1A} \\
	\delta Q_{r,U,{\rm B}} &= 2eU \int_{t_i}^{t_f} dt \int dt_1 \cdots dt_5 \int \frac{d\omega_1 d\omega_2}{(2\pi)^2} \sum_{r_1,r_2} \Gamma_{r} \Gamma_{r_1} \Gamma_{r_2} \nonumber \\
	&\hspace{-10mm}\times \mathrm{Re} \Bigl[\mathcal{G}_B(t_1,t_2,t_3,t_4,t_5)  f(\omega_1) f(\omega_2) e^{-i\omega_1(t_1-t_2)}e^{-i\omega_2(t_4-t_5)} \nonumber \\
	&\hspace{-10mm}\qquad \qquad \times C_{r_1}(\omega_1,t_1+t,t_2+t) C_{r_2}(\omega_2,t_4+t,t_5+t) \Bigr],
	\label{eqn:deltaQ1B}	
\end{align}
where $\mathcal{G}_A(t_1,t_2,t_3,t_4)$ and $\mathcal{G}_B(t_1,t_2,t_3,t_4,t_5)$ are give by
\begin{align}
	&\mathcal{G}_A(t_1,t_2,t_3,t_4) \nonumber \\ &= G_0^R(-t_2) G_0^R(t_2 - t_3) G_0^A(t_4 - t_2)G_0^R(t_2 - t_1), \\
	&\mathcal{G}_B(t_1,t_2,t_3,t_4,t_5) \nonumber \\ &= G_0^R(-t_3) G_0^R(t_3-t_4) G_0^A(t_5-t_3) G_0^R(t_3-t_1) G_0^A(t_2) \nonumber \\
	&+ G_0^R(-t_1) G_0^A(t_2-t_3) G_0^R(t_3-t_4) G_0^A(t_5-t_3) G_0^A(t_3).
\end{align}
Here the time variables are shifted to $t_n \to t_n +t$ to simplify the formula, and the abbreviation $G_0^{R(A)}(t_1-t_2) = G_{0,s}^{R(A)}(t_1,t_2)$ is introduced using the fact that the noninteracting retarded (advanced) GF is spin-independent and is not affected by external modulation.

\section{\label{noninteracting_part} Derivation of Eq.~(\ref{eqn:Q0pump})}
By applying the adiabatic approximation to Eqs.~(\ref{eqn:noninteracting_A}) and (\ref{eqn:noninteracting_B}), $\delta Q_{r,0,A}^{(1)}$ becomes
\begin{align}\label{eqn:adiabaticQ0A}
	&\delta Q_{r,0,A}^{\mathrm{pump}} = -4e \int_{t_i}^{t_f} dt \int dt_1\int \frac{d\omega}{2\pi} \nonumber \\
	&\times \mathrm{Re} \Bigl[ i\Gamma_r G_{0}^R(t,t_1) f(\omega) 
	e^{-i\omega (t_1-t)} C_r^{(1)}(\omega,t,t_1,0) \Bigr\}, \\
	\label{eqn:adiabaticQ0B}
	&\delta Q_{r,0,B}^{\mathrm{pump}} = 2e \sum_{r^{\prime}} \int_{t_i}^{t_f} dt \int dt_1dt_2 \int \frac{d\omega}{2\pi} \nonumber \\
	&\times \mathrm{Re} \Bigl[ \Gamma_r \Gamma_{r^{\prime}}G_{0}^R(-t_1) G_{0}^A(t_2) 
	f(\omega) e^{-i\omega(t_1-t_2)} C_{r^{\prime}}^{(1)}(\omega,t,t_1,t_2) \Bigr].
\end{align}
Substituting Eq.~(\ref{eqn:ad_c_1}) to (\ref{eqn:adiabaticQ0A}) and (\ref{eqn:adiabaticQ0B}), we obtain
\begin{align}
	\delta Q_{r,0,A}^{\mathrm{pump}} &=  e \sum_{\mu = 1}^{4} \int_{X^{\mu}(t_i)}^{X^{\mu}(t_f)} dX_{\mu} \int d\omega \Gamma_r  \frac{\partial \mathcal{A}(\omega)}{\partial \omega} \frac{\partial f_r(\omega)}{\partial X^{\mu}}, \\
	\delta Q_{r,0,B}^{\mathrm{pump}} &= -\pi e \sum_{\mu = 1}^{4} \int_{X^{\mu}(t_i)}^{X^{\mu}(t_f)} dX_{\mu} \int d\omega \Gamma_r \mathcal{A}^2(\omega) \frac{\partial f_{\mathrm{eff}}(\omega)}{\partial X^{\mu}}.
\end{align}
Here, $X^{\mu}$ is a parameter vector (Eq.~(\ref{eqn:param_vec})), $f_r(\omega)$ is the Fermi distribution function of reservoir $r$ (Eq.~(\ref{eq:distributionfunc})), $\mathcal{A}(\omega)$ is the spectral function (Eq.~(\ref{eqn:spectralfunction})), and $f_{\mathrm{eff}}(\omega)$ is the 
effective distribution function (Eq.~(\ref{eqn:EffectiveFermi})).
To obtain this formula, we use relations between derivatives of the Fermi distribution function such as,
\begin{align}
	\frac{\partial f_r(\omega)}{\partial \omega} = -\frac{\partial f_r(\omega)}{\partial \mu_r} = -\frac{T_r}{\omega - \mu_r} \frac{\partial f_r(\omega)}{\partial T_r},
	\label{eq:distribution}
\end{align}
and have converted integral variable as
\begin{align}
	\int_{t_i}^{t_f} dt \dot{X}_{\mu}(t) (\cdots) = \int_{X^{\mu}(t_i)}^{X^{\mu}(t_f)} dX_{\mu} (\cdots). 
\end{align}
Summation of $\delta Q_{r,0,A}^{\mathrm{pump}}$ and $\delta Q_{r,0,B}^{\mathrm{pump}}$ gives $\delta Q_{r,0}^{\mathrm{pump}}$ as
\begin{align}
	\delta Q_{r,0}^{\mathrm{pump}} & = \delta Q_{r,0,A}^{\mathrm{pump}} + \delta Q_{r,0,B}^{\mathrm{pump}} \nonumber \\
	&= \sum_{\mu = 1}^{4} \int_{X_{\mu}(t_i)}^{X_{\mu}(t_f)} \! \! \! dX_{\mu} \pi_{r,0}^{\mu}(\bm{X}),  \\
	\pi_{r,0}^{\mu}(\bm{X}) &= \frac{\partial}{\partial X_\mu} \biggl( 
	 e  \int d\omega \Gamma_r 
	 \bigl[ (\partial_{\omega} \mathcal{A}(\omega)) f_r(\omega; \bm{X}) \Biggr. \nonumber \\
	& \hspace{10mm} \biggl.
	- \pi \mathcal{A}^2(\omega) f_{\mathrm{eff}}(\omega; \bm{X}) \bigr]\biggr) .
\end{align}

\section{\label{rep_of_pi}Derivation of Eqs.~(\ref{eqn:gene_pump_c})-(\ref{eqn:pi_only_pump})}
By applying the adiabatic approximation to Eqs.~(\ref{eqn:deltaQ1A}) and (\ref{eqn:deltaQ1B}), we obtain
\begin{align}\label{eqn:deltaQ1A_adiabatic}
	& \delta Q_{r,U,A}^{\mathrm{pump}} =   -4eU \sum_{r^{\prime}} \int \frac{d\omega_1 d\omega_2}{(2\pi)^2} \int_{t_i}^{t_f} dt \int dt_1 \cdots dt_4  \Gamma_{r^{\prime}} \Gamma_{r} \nonumber \\
	&\times \mathrm{Re} \Bigl[ i \mathcal{G}_A(t_1,t_2,t_3,t_4) f(\omega_1) f(\omega_2) e^{-i\omega_1 t_1}e^{-i\omega_2(t_3-t_4)}   \nonumber \\
	& \qquad \qquad \times \Bigl( C_{r}^{(1)}(\omega_1,t,t_1,t) C_{r^{\prime}}^{(0)}(\omega_2,t,t_3,t_4) \nonumber \\
	& \qquad \qquad \quad + C_{r}^{(0)}(\omega_1,t,t_1,0) C_{r^{\prime}}^{(1)}(\omega_2,t,t_3,t_4) \Bigr) \Bigr],\\
\label{eqn:deltaQ1B_adiabatic}
	&\delta Q_{r,U,B}^{\mathrm{pump}} = 2eU \int_{t_i}^{t_f} dt \int dt_1 \cdots dt_5 \int \frac{d\omega_1 d\omega_2}{(2\pi)^2} \sum_{r_1,r_2} \Gamma_{r} \Gamma_{r_1} \Gamma_{r_2}  \nonumber \\
	&\times \mathrm{Re} \Bigl[\mathcal{G}_B(t_1,t_2,t_3,t_4,t_5)  f(\omega_1) f(\omega_2) e^{-i\omega_1(t_1-t_2)}e^{-i\omega_2(t_4-t_5)} \nonumber \\
	&\qquad \qquad \times \Bigl( C_{r_1}^{(1)}(\omega_1,t,t_1,t_2) C_{r_2}^{(0)}(\omega_2,t,t_4,t_5) \nonumber \\
	&\qquad \qquad \quad + C_{r_1}^{(0)}(\omega_1,t,t_1,t_2) C_{r_2}^{(1)}(\omega_2,t,t_4,t_5) \Bigr) \Bigr] .
\end{align}
Adding $\delta Q_{r,U,A}^{\mathrm{pump}}$ and $\delta Q_{r,U,B}^{\mathrm{pump}}$ and substituting Eqs.~(\ref{eqn:ad_c_0}) and (\ref{eqn:ad_c_1}) to Eqs.~(\ref{eqn:deltaQ1A_adiabatic}) and (\ref{eqn:deltaQ1B_adiabatic}), the expression of $\delta Q_{r,U}^{\mathrm{pump}}$ given in Eqs.~(\ref{eqn:gene_pump_c})-(\ref{eqn:pi_only_pump}) is derived.

\section{Circuit elements in AC transport} \label{App:DynamicalResistance}

This Appendix describes how the time delay $\delta t$ given in Eq.~(\ref{eq:deltat}) is written in terms of effective circuit elements.
The complex admittance of a coherent conductor coupled to the reservoir $r$ is calculated by using the linear response theory as ~\cite{Buttiker93b,Nigg06,Gabelli06,Gabelli07}
\begin{align}
G_r(\omega) &= \frac{I_r(\omega)}{\mu_r(\omega)} 
= - i C_r \omega+ R_r C_r^2 \omega^2 + O(\omega^3) \nonumber \\
&= \frac{1}{R_r -1/(i C_r \omega)} + O(\omega^3).
\end{align}
Here, $C_r$ and $R_r$ are circuit elements denoted as the dynamic capacitance and the dynamic resistance, respectively.
For the QD system, they are given by
\begin{align}
C_r &= e^2 \int d\omega
\mathcal{A}(\omega) 
\left(-\frac{\partial f_r}{\partial \omega} \right), \\
R_r &= \frac{\pi}{e^2} \frac{\displaystyle{\int d\omega \mathcal{A}^2(\omega) \left( -\frac{\partial f_r}{\partial \omega}\right)}}
{\displaystyle{\left(\int d\omega \mathcal{A} (\omega) \left(-\frac{\partial f_r}{\partial \omega} \right)\right)^2}}.
\end{align}
Recently, circuit elements are calculated for coherent transport under AC temperature modulation~\cite{Lim13}:
\begin{align}
G_{r,T}(\omega) &= \frac{I_r(\omega)}{T_r(\omega)}  = - i C_{r,T} \omega+ R_{r,T} C_{r,T}^2 \omega^2 + O(\omega^3) \nonumber \\
&= \frac{1}{R_{r,T} -1/(i C_{r,T} \omega)} + O(\omega^3), \\
C_{r,T} &= e^2 \int d\omega \mathcal{A}(\omega) \frac{\omega - \mu_r}{T_r}
\left(-\frac{\partial f_r}{\partial \omega}\right), \\
R_{r,T} &= \frac{\pi}{e^2}
\frac{\displaystyle{\int d\omega
\mathcal{A}^2(\omega) \frac{\omega - \mu_r}{T_r}  \left( -\frac{\partial f_r}{\partial \omega}\right)}}
{\displaystyle{\left(\int d\omega 
\mathcal{A}(\omega)\frac{\omega - \mu_r}{T_r} \left(-\frac{\partial f_r}{\partial \omega}\right) \right)^2}}.
\end{align}
Using Eq.~(\ref{eq:distribution}), $\delta t$ is rewritten by these circuit elements as
\begin{align}
\delta t = \frac{ \displaystyle{\sum_{r=L,R} \left(R_r C_r^2  \dot{\mu}_r+ R_{r,T} C_{r,T}^2 \dot{T}_r \right)}}
{\displaystyle{\sum_{r=L,R} \left(C_r \dot{\mu}_r+ C_{r,T} \dot{T}_r \right)}}.
\end{align}
This relation implies that the time delay $\delta t$ reflect a relaxation time of coherent transport via dynamical resistances and capacitances~\cite{footnoteAC}.

\bibliography{apssamp}

\begin{thebibliography}{99}
\bibitem{footnote:tmfield}
We note that in Ref.~\citen{Eich14}, there is a mistake in introducing
a gauge potential and a thermomechanical field.
\bibitem{footnote:discussion}
In Ref.~\citen{Altimiras10}, edge states of integer quantum Hall effect have been used.
As the edge states of this system are spin-polarized, we need to introduce Coulomb interaction by, e.g., using two QDs made from the edge states.
\bibitem{footnoteAC}
In mesoscopic devices in a system composed of one lead and one dot, dynamic response is understood by a series circuit of the dynamic resistance and the dynamical capacitance. The delay time for the present model is, however, expressed by a complex linear combination of circuit elements, and cannot be interpreted by a simple series circuit of these circuit elements.
\end{thebibliography}

\end{document}